# Active Estimation of Multiplicative Faults in Dynamical Systems

Gabriel de A. Gleizer[a], Peyman Mohajerin Esfahani[a,b], Tamas Keviczky[a]

June 30, 2025

**Abstract**

This paper addresses the problem of estimating multiplicative fault signals in linear time-invariant systems by processing its input and output variables, as well as designing an input signal to maximize the accuracy of such estimates. The proposed real-time fault estimator is based on a residual generator used for fault detection and a multiple-output regressor generator, which feed a moving-horizon linear regression that estimates the parameter changes. Asymptotic performance guarantees are provided in the presence of noise. Motivated by the performance bounds, an optimal input design problem is formulated, for which we provide efficient algorithms and optimality bounds. Numerical examples demonstrate the efficacy of our approach and the importance of the optimal input design for accurate fault estimation.

## 1 Introduction

In control applications, dynamical systems can occasionally display unexpected behavior due to the presence of faults. Faults can occur due to many reasons, such as unexpected disturbances or malicious attacks, failure of some component, or wear and tear. If not detected timely, faults can lead to economic losses due to out-of-specification performance and system shutdown, or even catastrophic events incurring human, property, and/or environmental damage. For this reason, fault diagnosis methods have been subject of intense research since the early seventies [1], [2] and have since been applied to a myriad of applications, such as robots, transport systems, power systems, manufacturing processes, and chemical processes [3]. Particularly for high-performance control systems, such as high-tech machines, not only the detection of faulty behavior is important, but also the estimation of how much change has occurred, a problem called *fault estimation,* which can be further classified in two types of faults: *additive* and *multiplicative*. This distinction is made depending on the end goal of the technological health monitoring system. If it is sufficient to model the fault in question as an external signal that enters additively in the system equations, one has an *additive fault estimation* problem; in this case, one is satisfied with tracking such an external signal. Instead, if it is more appropriate to model the fault as a change in system parameters, or additional intrusive dynamics, it becomes a problem of *multiplicative fault estimation*. The latter is the topic of the present paper. When faults are additive and the system is linear, the estimation performance generally does not depend on the input to the system, and hence active fault estimation is unnecessary. In contrast, this separation property does not hold for multiplicative faults. For example, changes in friction components of a mechanical system can only be detected when it moves. Therefore, the ability of the input signals to properly excite the system is a critical component of multiplicative fault estimation. When a fault estimation scheme involves input design or control design, it is called *active fault estimation*. Given its impact in the estimation performance, input design is also a central part of this paper.

The idea of designing inputs to increase the diagnosability of faults was probably pioneered by [4] using a statistical framework. Later, a deterministic multi-model approach was proposed by [5], whose objective is to determine which model generates the observed trajectory between two candidates. This has been extended to

This work was supported by the Digital Twin project with project number P18-03 of the research programme TTW-Perspectief, which is partly financed by the Dutch Research Council (NWO). Peyman Mohajerin Esfahani acknowledges the support of the ERC grant TRUST949796. The authors are with (a) Delft University of Technology, The Netherlands; and (b) University of Toronto, Canada.

multiple model candidates in a hybrid stochastic-deterministic framework in [6]. As an alternative to offline design of input signals, [7] considers a closed loop approach where the residual generator also creates a signal to be fed back to the system in order to improve fault detectability and/or tolerance. This problem is still an active area of research [8], [9], [10]. An overview of active fault detection can be found in [11], and of input design for fault diagnosis in [12].

Most of the focus of the literature on active diagnosis has been on classification problems as reported above, as opposed to fault *estimation*. Nonetheless, the related field of experiment design for system identification is much more well established. It emerged in the seventies (see, e.g., [13] [14], [15]) inspired by experiment design in the statistics literature. The idea is to design an input that maximizes a function of the Fischer Information Matrix (FIM), which depends on input data, based on the observation that the variance of a linear estimator is inversely proportional to it. This leads to so-called *T-optimal* and *E-optimal* designs, which maximize, respectively, the trace and the minimum eigenvalue of the FIM. After that, most focus has been given to problems associated to identification for control (e.g., [16]), closed-loop identification (e.g., see [17], which also serves as an overview of this literature), and frequency domain identification, see, e.g., [18], [19]. More recently, [20] addressed the problem of computing the T-optimal input in the MIMO case where the total energy of the input and output are bounded. They provide global optimal results for the case of only two quadratic constraints building on the results of [21]. With the recent surge of data-driven control, experiment design has also been gaining increased attention [22], [23], [24]. In [22], a series of experiments is designed to achieve the most accurate system identification with the least samples, using the classical FIM-optimal approaches as an oracle. In [23], [24], the focus is qualitative rather than quantitative: the goal is to derive sufficient excitation conditions that enable designing a stabilizing controller from an experiment. Since our focus is quantitative, our experiment design problem is closer to that of [13], [14], [15],[16],[17], [20] and related work.

In this work, we propose a fault estimation method for LTI systems described by differential-algebraic equations (DAEs), which include classical ordinary differential and difference equations as a special case [25], [26], [27]. The method is designed to estimate *multiplicative* faults. Additionally, we propose an optimal input design algorithm that maximizes the accuracy of the aforementioned method. The main contributions are summarized in the following list:

1. **Estimation of multiplicative faults affecting latent variables.** We extend the work in [28] to multiple simultaneous faults that possibly multiply latent variables such as states and external disturbances. The fault estimation system has the same structure as [28] (see Fig. 1), including a residual generator, a regressor generator, and a fault estimator. Due to the fault characteristics mentioned above, a novel formulation for the design of residual and regressor generators is required, which we provide in Theorem 3.1. The fault estimation is obtained with a Gauss–Newton-like approach.

2. **Performance characterization: bias and variance.** We characterize the bias and variance bounds of the least-squares fault estimation system in the presence of measurement noise (Theorem 3.2), assuming the regressor signal is persistently exciting. The bias is zero in the special case where the multiplicative faults affect only the measured variables (Corollary 3.1).

3. **Optimal input design.** Motivated by the error characterization formula of Theorem 3.2, we propose an E-optimal periodic input design that maximizes the asymptotic value of the regressor's singular value in the absence of external disturbances (Definition 4.1).

    **3a. Exact formulation and first-order information.** We provide an exact formulation of the optimal periodic input design problem, including exact first-order information via a subgradient (Theorem 4.1). This paves the way to a local optimization method based on convex optimization that yields fast and efficient local optima (Algorithm 1).

    **3b. Convex relaxation and sub-optimality gap.** We provide a convex semi-definite relaxation (Proposition 4.2), which gives a conservative certificate of the local solution's sub-optimality gap.

This paper is organized as follows. Section 2 provides mathematical preliminaries and presents the problem formulation. Section 3 presents the fault estimator design and provides formal results on the estimation performance. Section 4 presents the input design problem and methods to solve it. Section 5 shows numerical simulations, highlighting how input design plays a major role in the estimation accuracy. Section 6 concludes the paper with discussions and ideas for future work. Finally, Section 7 contains the technical proofs, which we set apart from the main text for improved readability.

**Mathematical notation.** Throughout the text, we use Italic typesetting, e.g., $a$, for scalars or scalar-valued functions; bold letters, e.g., $\boldsymbol{a}$, for vectors or vector-valued functions; bold capital letters, e.g., $\boldsymbol{A}$, for matrices; and calligraphic letters, e.g., $\mathscr{A}$, for sets or operators in Hilbert spaces such as transfer functions. We denote by $\mathbb{N}_0$ the set of natural numbers including zero, $\mathbb{N} := \mathbb{N}_0 \setminus \{0\}$, $\mathbb{N}_{\leq n} := \{1, 2, ..., n\}$, and by $\mathbb{R}_+$ the set of non-negative reals. We denote by $|\boldsymbol{x}|_p$ the $p$-norm of the vector $\boldsymbol{x} \in \mathbb{R}^n$, dropping the subscript when $p = 2$. We denote by $\|\boldsymbol{A}\|_p$ the $p$-induced norm of the matrix $\boldsymbol{A} \in \mathbb{R}^{n \times m}$. For a square matrix $\boldsymbol{A} \in \mathbb{R}^{n \times n}$, $\lambda(\boldsymbol{A}) \subset \mathbb{C}^n$ is the set of its eigenvalues, and $\lambda_i(\boldsymbol{A})$ is the $i$-th largest-in-magnitude. For any matrix $\boldsymbol{A} \in \mathbb{R}^{m \times n}$, $s(\boldsymbol{A}) \subset \mathbb{R}_+$ denotes the set of its singular values, with $s_i$ being the $i$-th largest. The complex conjugate of $z \in \mathbb{C}^n$ is denoted by $z^*$. The set $\mathbb{S}^n$ denotes the set of symmetric matrices in $\mathbb{R}^n$. For $\boldsymbol{P} \in \mathbb{S}^n$, we write $\boldsymbol{P} \succ \boldsymbol{0}$ ($\boldsymbol{P} \succeq \boldsymbol{0}$) if $\boldsymbol{P}$ is positive definite (semi-definite); and $\boldsymbol{P} \prec \boldsymbol{0}$ ($\boldsymbol{P} \preceq \boldsymbol{0}$) if $\boldsymbol{P}$ is negative definite (semi-definite). The set of positive (semi-)definite matrices in $\mathbb{S}^n$ is denoted by $\mathbb{S}^n_{++}$ ($\mathbb{S}^n_{+}$). The partial order induced by $\preceq$ allows us to use $\boldsymbol{P} \preceq \boldsymbol{S}$ when $\boldsymbol{P} - \boldsymbol{S} \preceq \boldsymbol{0}$ (and equivalently for $\boldsymbol{P} \succeq \boldsymbol{S}$).

For a signal $\boldsymbol{x} : \mathscr{T} \to \mathbb{R}^n$, where $\mathscr{T}$ is the time axis (typically $\mathbb{N}_0$ or $\mathbb{R}_+$), we denote by $\boldsymbol{X}^h_{[a,a+(N-1)h]}$ the $N$ by $n$ matrix $[\boldsymbol{x}(a) \quad \boldsymbol{x}(a+h) \quad \cdots \quad \boldsymbol{x}(a+(N-1)h)]$. When the sampling interval $h$ and the time range $[a, a+(N-1)h]$ are clear from context, we may drop the subscript and superscript, and the mere capitalization shall indicate the transformation.

# 2 Preliminaries and problem formulation

## 2.1 System model

Throughout this work, we consider faulty linear time-invariant dynamical systems described in a differential-algebraic equation (DAE) form,

$$\left(\boldsymbol{H}(\mathfrak{q}) + \sum_{i=1}^{m} f_i \boldsymbol{H}'_i(\mathfrak{q})\right)\boldsymbol{\xi} + \left(\boldsymbol{L}(\mathfrak{q}) + \sum_{i=1}^{m} f_i \boldsymbol{L}'_i(\mathfrak{q})\right)\boldsymbol{z} + \boldsymbol{W}(\mathfrak{q})\boldsymbol{w} = 0, \tag{1}$$

where $\boldsymbol{\xi}, \boldsymbol{z}$ and $\boldsymbol{w}$ are (either continuous-time or discrete-time) signals taking values in $\mathbb{R}^{n_\xi}$, $\mathbb{R}^{n_z}$ and $\mathbb{R}^{n_w}$, respectively. The matrices $\boldsymbol{H}(\mathfrak{q}), \boldsymbol{L}(\mathfrak{q}), \boldsymbol{W}(\mathfrak{q}), \boldsymbol{H}'_i(\mathfrak{q}), \boldsymbol{L}'_i(\mathfrak{q})$, for $i \in \mathbb{N}_{\leq m}$ are polynomial matrices in the operator $\mathfrak{q}$, with $n_r$ rows and compatible number of columns. The operator $\mathfrak{q}$ is a linear operator in the signal space where $\boldsymbol{\xi}, \boldsymbol{z}$ and $\boldsymbol{w}$ are defined, typically the forward shift operator for discrete-time signals or the (right) time derivative for continuous-time (right) differentiable signals. The scalars $f_i$ represent unknown multiplicative faults, which are assumed to be constant for analysis; in Section 3.3 we show how to extend the framework for time-varying faults. The signal $\boldsymbol{\xi}$ comprises the system's latent variables, such as internal states and external disturbances, $\boldsymbol{z}$ comprises known signals such as control inputs, measured outputs and references, and $\boldsymbol{w}$ represents unknown i.i.d. noise with zero mean and variance $\sigma^2$. All components of $\boldsymbol{w}$ can have the same variance without loss of generality, by rescaling the rows of (1). One may notice that we have partitioned unknown variables in two; this is a common approach in the fault diagnosis literature, that allows one to distinguish unknown variables $\boldsymbol{\xi}$ that do not have an a priori statistical description, and are determined either by dynamics or an external agent, from unknown variables $\boldsymbol{w}$ which may be characterized as random processes. We make the following nominal (disturbance) observability assumption.

**Assumption 2.1. (Nominal observability)** The polynomial matrix $\boldsymbol{H}(\mathfrak{q})$ has full column rank for all $\mathfrak{q} \in \mathbb{C}$. Equivalently, $\boldsymbol{H}(\mathfrak{q})$ has a polynomial left-inverse $\boldsymbol{H}(\mathfrak{q})^\dagger$ such that $\boldsymbol{H}(\mathfrak{q})^\dagger \boldsymbol{H}(\mathfrak{q}) = \mathbf{I}$ (see, e.g., [29]).

We call Assumption 2.1 a nominal observability condition because, when $f_i = 0$ for all $i$ and $\boldsymbol{w} \equiv \boldsymbol{0}$, the unmeasured variables $\boldsymbol{\xi}$ satisfy $\boldsymbol{\xi} = -\boldsymbol{H}(\mathfrak{q})^\dagger \boldsymbol{L}(q)\boldsymbol{z}$; that is, they can be determined by combinations of differentials of the measured variables $\boldsymbol{z}$. Note that observability is only required in the nominal case, i.e., the faults $f_i$ can be estimated even if $\boldsymbol{H}(\mathfrak{q}) + \sum_{i=1}^m f_i \boldsymbol{H}'_i(\mathfrak{q})$ loses rank. One advantage of the DAE framework used here is that, even though we impose an observability condition, it does not require building (unknown input) observers to obtain a fault estimation system, which has great scalability implications.

**Remark 2.1. (Converting state-space ODE to DAE)** The DAE framework in (1) encompasses standard differential–algebraic state-space LTI formulations of the form

$$\left(G+\sum_{i=1}^{m} f_i G_i'\right)x(k+1)=\left(A+\sum_{i=1}^{m} f_i A_i'\right)x(k)+\left(B_u+\sum_{i=1}^{m} f_i B_i'\right)u(k)+\left(B_d+\sum_{i=1}^{m} f_i B_{d,i}'\right)d(k)+B_w w(k),$$
$$y(k)=\left(C+\sum_{i=1}^{m} f_i C_i'\right)x(k)+\left(D_u+\sum_{i=1}^{m} f_i D_i'\right)u(k)+\left(D_d+\sum_{i=1}^{m} f_i D_{d,i}'\right)d(k)+D_w w(k), \tag{2}$$

where $x : \mathbb{N}_0 \to \mathbb{R}^{n_x}$ is the state trajectory, $u : \mathbb{N}_0 \to \mathbb{R}^{n_u}$ is the input signal, $y : \mathbb{N}_0 \to \mathbb{R}^{n_y}$ is the output signal, and $d : \mathbb{N}_0 \to \mathbb{R}^{n_d}$ is the external disturbance signal. By taking $z = \begin{bmatrix} y^\top & u^\top \end{bmatrix}^\top$ and $\xi = \begin{bmatrix} x^\top & d^\top \end{bmatrix}^\top$, the following conversion from (2) to (1) holds: for all $i = 1, ..., m$,

$$H(\mathrm{q}) = \begin{bmatrix} -\mathrm{q}G + A & B_d \\ C & D_d \end{bmatrix}, \qquad L(\mathrm{q}) = \begin{bmatrix} 0 & B_u \\ -I & D_u \end{bmatrix}, \qquad W(\mathrm{q}) = \begin{bmatrix} B_w \\ D_w \end{bmatrix},$$
$$H_i'(\mathrm{q}) = \begin{bmatrix} -\mathrm{q}G_i' + A_i' & B_{d,i}' \\ C_i' & D_{d,i}' \end{bmatrix}, \qquad L_i'(\mathrm{q}) = \begin{bmatrix} 0 & B_i' \\ 0 & D_i' \end{bmatrix}.$$

## 2.2 Linear-algebra operations in polynomial matrices

It is possible translate certain polynomial matrix equations into standard linear algebraic ones, which are suitable for efficient numerical methods from linear algebra and optimization. These techniques have been used in [26] and others. For that, let us introduce some notation: given a polynomial matrix $H(\mathrm{q}) := \sum_{i=0}^{d} H_i \mathrm{q}^i$ of degree $d$, we denote

$$\mathrm{blkrow}(H(\mathrm{q})) := \begin{bmatrix} H_0 & H_1 & \cdots H_d \end{bmatrix},$$
$$\bar{H} := \begin{bmatrix} H_0 & \cdots & H_d & 0 & \cdots & 0 \\ 0 & H_0 & \cdots & H_d & 0 & \cdots \\ \vdots & & \ddots & & \ddots & \\ 0 & \cdots & 0 & H_0 & \cdots & H_d \end{bmatrix},$$

where blkrow($H(\mathrm{q})$) is the block-row form of $H(\mathrm{q})$ and $\bar{H}$ is its block-Toeplitz form. In the latter, the number of block rows shall be derived from the context of its usage, which we provide next:

**Lemma 2.1. (Characterization of polynomial matrix multiplication [27, Lemma 4.2])** *Let $H_1(\mathrm{q})$ and $H_2(\mathrm{q})$ be polynomial matrices of degrees $d_1$ and $d_2$, respectively, and with dimensions allowing the product $H_1(\mathrm{q})H_2(\mathrm{q})$. The following identity holds:*

$$\mathrm{blkrow}(H_1(\mathrm{q})H_2(\mathrm{q})) = \mathrm{blkrow}(H_1(\mathrm{q}))\bar{H}_2,$$

*where the number of block rows of $\bar{H}_2$ is $d_1 + 1$.*

## 2.3 Problem statement

In this work we are interested in three problems: (i) to design a fault estimation filter, (ii) to characterize its performance bounds, and (iii) to design an input signal within given constraints that minimizes the fault estimation error. We formalize these problems below.

**Problem 2.1 (Estimator design).** Consider system (1), and recall that the signal $z$ is available. We aim to design a (nonlinear) fault estimation filter $\mathscr{F}$ that generates an estimate *signal* $\hat{f}$ for the multiplicative fault $f := [f_1 \quad f_2 \quad \cdots \quad f_m]^\top$ such that

1. **Causality:** the filter $\mathscr{F}$ is a causal dynamical system taking $z$ as input;
2. **Estimation error order:** the multiplicative estimation error $|\hat{f}(t) - f|$ in the absence of noise is on the order of $|f|^2$;

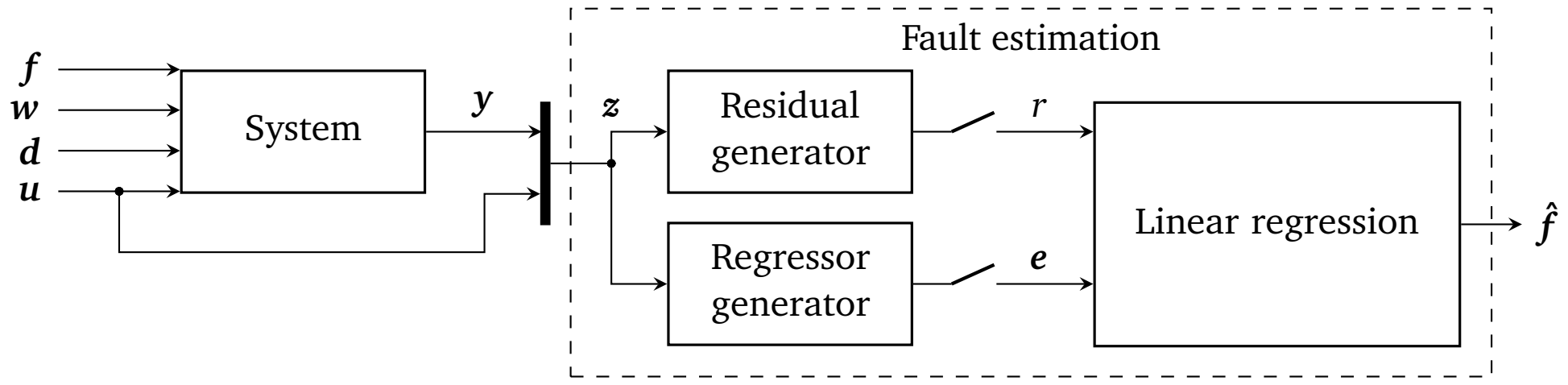

Figure 1: The proposed fault estimation architecture. The switches represent samplers, since the linear regression always operates in discrete-time. The system, residual generator and regressor generator can be either continuous- or discrete-time.

3. **Asymptotic consistency:** for sufficiently high signal-to-noise ratio, the *asymptotic expected squared error* $\lim_{t\to\infty} \mathbb{E}[|\hat{\boldsymbol{f}}(t) - \boldsymbol{f}|^2]$ is on the order of $\sigma^2(1 + |\boldsymbol{f}|^2)$, with the bias on the order of $|\boldsymbol{f}|^2$.

The estimation error specifications in Problem 2.1 depend on $|\boldsymbol{f}|^2$, making them suitable for small faults $\boldsymbol{f}$.

The second task is characterizing performance bounds of our designed fault estimation method. Again, focusing on the small-fault scenario, we focus on asymptotic results as $\boldsymbol{f}$ approaches zero. Following the stochastic perturbation theory of [30], we concentrate our attention to the *first-order approximation* $\check{\boldsymbol{f}}$ of the estimate $\hat{\boldsymbol{f}}$.

**Problem 2.2** (**Performance bounds**)**.** Given system (1) and the designed fault estimator filter $\mathscr{F}$, determine bounds on the asymptotic bias and variance of the first-order approximation of the error $\check{\boldsymbol{f}}(t) - \boldsymbol{f}(t)$.

Designing a good fault estimation scheme for multiplicative faults may not be sufficient for their accurate estimation. Effective excitation of the system's states is also necessary. Therefore, for active estimation of the faults $\boldsymbol{f}$, we pose the following experiment design problem:

**Problem 2.3** (**Input design**)**.** Given system (1) and the designed fault estimator filter $\mathscr{F}$, compute a periodic signal $\boldsymbol{u}$ that minimizes the asymptotic performance bounds obtained by solving Problem 2.2.

**Remark 2.2. (Finite-time vs. asymptotic performance)** Asymptotic performance criteria measure performance after the contribution of the initial state vanishes. In this case, input periodicity ensures that the performance metric $\mathbb{E}[|\hat{\boldsymbol{f}}(t) - \boldsymbol{f}(t)|^2]$ converges as $t$ grows large. If the initial state *is known*, it can be easily incorporated in the input design problem, and a $T$-periodic input renders $\mathbb{E}[|\hat{\boldsymbol{f}}(t) - \boldsymbol{f}(t)|^2]$ constant for all $t \geq T$.

# 3 Fault estimation of multiple multiplicative faults

Inspired by [28], we design a system composed of three blocks as depicted in Fig. 1. One block is a residual generator, which as usual should have zero response if no faults are present, but we shall design to have specific sensitivity properties to the faults $f_i$ that allows them to be distinguishable among each other. The block below it is a regressor generator, similar to what is called pre-filter in [28], whose function is to provide a basis of signals that serves as regressors to the final block, the fault estimator, which in turn performs a linear regression in a moving-horizon fashion.

Both the residual and the regressor generators are LTI systems. They shall be designed, respectively, as the transfer functions $d(\mathrm{q})^{-1}\boldsymbol{N}(\mathrm{q})$ and $d(\mathrm{q})^{-1}\boldsymbol{M}(\mathrm{q})$ with $n_\mathrm{z}$ inputs and 1 and $m$ outputs, respectively, where $d(\mathrm{q})$ is a polynomial designed to be stable, provide good noise attenuation, and render the generators proper. As a first requirement, the residual $r$ must satisfy, under small fault assumptions, the approximation

$$r \approx \boldsymbol{f}^\mathrm{T} d(\mathrm{q})^{-1}\boldsymbol{M}(\mathrm{q})\boldsymbol{z}. \tag{3}$$

That is, the residual is approximately the linear combination of the regressors. This paves the way for applying linear regression to obtain an estimate $\hat{f}_i(t)$ of each fault $f_i$, from a sampled trajectory of $r$ and $d(\mathrm{q})^{-1}\boldsymbol{M}(\mathrm{q})\boldsymbol{z}$ from $t - T$ to $t$.

This framework imposes certain requirements to the residual and regressor generators. In addition to (3), we want that the term $\boldsymbol{f}^\mathsf{T} d(\mathrm{q})^{-1}\boldsymbol{M}(\mathrm{q})\boldsymbol{z}$ is not identically zero for nonzero faults $\boldsymbol{f}$, and that $\boldsymbol{M}(\mathrm{q})\boldsymbol{z}$ spans linearly independent signals for sufficiently exciting $\boldsymbol{z}$. For the excitation condition, we invoke a recent notion originally proposed for continuous-time systems:

**Definition 3.1. (Persistency of excitation [31, Def. 1])** A signal $\boldsymbol{z} : \mathcal{T} \to \mathbb{R}^m$ is *persistently exciting (PE) of order $d$* if, for all $\boldsymbol{v} \in \mathbb{R}^{md}$,

$$\boldsymbol{v}^\mathsf{T} \begin{bmatrix} \boldsymbol{z}(t) \\ \mathrm{q}\boldsymbol{z}(t) \\ \vdots \\ \mathrm{q}^{d-1}\boldsymbol{z}(t) \end{bmatrix} = 0 \ \forall t \implies \boldsymbol{v} = \boldsymbol{0}. \tag{4}$$

The design requirements are summarized in the definition below:

**Definition 3.2. (Independently sensitive residual generator)** Consider system (1) with noise $\boldsymbol{w} \equiv \boldsymbol{0}$. A BIBO stable linear filter $d^{-1}(\mathrm{q})\boldsymbol{N}(\mathrm{q})\boldsymbol{L}(\mathrm{q})$ is called an *independently sensitive residual generator* if there exist a polynomial matrix $\boldsymbol{M} \in \mathbb{R}[\mathrm{q}]^{m \times n_z}$ such that the following conditions hold:

1. $\lim_{t\to\infty} r(t) = 0 \iff \boldsymbol{f} \equiv \boldsymbol{0}$, i.e., the signal $r$ is a fault detection residual;
2. for any norm $|\cdot|$, $\lim_{|f|\to 0} r - \boldsymbol{f}^\mathsf{T} d(\mathrm{q})^{-1}\boldsymbol{M}(\mathrm{q})\boldsymbol{z} = 0$, i.e., the approximation (3) holds;
3. if $\boldsymbol{z}$ is PE of sufficiently high order, the signals $d(\mathrm{q})^{-1}\boldsymbol{M}(\mathrm{q})\boldsymbol{z}$ are linearly independent.

The corresponding filter $d(\mathrm{q})^{-1}\boldsymbol{M}(\mathrm{q})$ is called a *regressor generator.*

## 3.1 Residual and regressor generator design

Consider a standard residual generator used for the additive fault case, i.e., one taking the form $d(\mathrm{q})r = \boldsymbol{N}(\mathrm{q})\boldsymbol{L}(\mathrm{q})\boldsymbol{z}$, where $\boldsymbol{N}(\mathrm{q})\boldsymbol{H}(\mathrm{q}) \equiv \boldsymbol{0}$. The first observation is that, when $f_i \not\equiv 0$, $\boldsymbol{N}(\mathrm{q})$ generally ceases to be in the left null space of the system (1). I.e., $\boldsymbol{N}(\mathrm{q})\boldsymbol{H}(\mathrm{q}) \equiv 0$ does not imply that $\boldsymbol{N}(\mathrm{q})(\boldsymbol{H}(\mathrm{q}) + \sum_i f_i \boldsymbol{H}_i'(\mathrm{q})) \equiv 0$. In this case, the residual output is affected by input–output data of the system $\boldsymbol{z}$ as well as the latent variables $\boldsymbol{\xi}$. However, this dependency is largely predictable if the faults $f_i$ are sufficiently small, as we show next.

**Lemma 3.1 (Residual characterization).** *Consider system* (1) *and a residual generator described by $d(\mathrm{q})r = \boldsymbol{N}(\mathrm{q})\boldsymbol{L}(\mathrm{q})\boldsymbol{z}$ satisfying $\boldsymbol{N}(\mathrm{q})\boldsymbol{H}(\mathrm{q}) \equiv \boldsymbol{0}$. If Assumption 2.1 holds, then the residual $r$ satisfies*

$$d(\mathrm{q})r = \boldsymbol{N}(\mathrm{q})\sum_i f_i \boldsymbol{G}_i(\mathrm{q})\boldsymbol{z} + \sum_{i,j} f_i f_j \boldsymbol{J}_{i,j}(\mathrm{q}) \begin{bmatrix} \boldsymbol{\xi} \\ \boldsymbol{z} \end{bmatrix} + \sum_i f_i \boldsymbol{F}_i(\mathrm{q})\boldsymbol{w} - \boldsymbol{N}(\mathrm{q})\boldsymbol{W}(\mathrm{q})\boldsymbol{w}. \tag{5}$$

*where $\boldsymbol{G}_i(\mathrm{q}) := \boldsymbol{H}_i'(\mathrm{q})\boldsymbol{H}(\mathrm{q})^\dagger \boldsymbol{L}(\mathrm{q}) - \boldsymbol{L}_i'(\mathrm{q})$ for all $i = 1, ..., m$, $\boldsymbol{H}^\dagger(\mathrm{q})$ is a polynomial matrix satisfying $\boldsymbol{H}^\dagger(\mathrm{q})\boldsymbol{H}(\mathrm{q}) = \boldsymbol{I}$ and $\boldsymbol{J}_{i,j}(\mathrm{q})$ and $\boldsymbol{F}_i(\mathrm{q})$ are polynomial matrices. Furthermore, if $\boldsymbol{H}_i'(\mathrm{q}) \equiv \boldsymbol{0}$ for all $i = 1, ..., m$, then $\boldsymbol{J}_{i,j}(\mathrm{q}) \equiv \boldsymbol{0}$ and $\boldsymbol{F}_i(\mathrm{q}) \equiv \boldsymbol{0}$ for all $i, j = 1, ..., m$.*

*Proof:* See Section 7.1. ■

**Remark 3.1 (Linear-algebraic computation of the left-inverse of a polynomial matrix).** From Assumption 2.1, there exists a polynomial matrix $\boldsymbol{H}(\mathrm{q})^\dagger$ satisfying $\boldsymbol{H}(\mathrm{q})^\dagger \boldsymbol{H}(\mathrm{q}) = \mathbf{I}$. Hence, for a sufficiently large natural number $k$, there exist $\boldsymbol{H}_i^\dagger, i = 0, \ldots, k$ such that

$$\boldsymbol{H}^\dagger(\mathrm{q})\boldsymbol{H}(\mathrm{q}) = \begin{bmatrix} \boldsymbol{H}_0^\dagger & \cdots \boldsymbol{H}_k^\dagger \end{bmatrix} \begin{bmatrix} \boldsymbol{H}_0 & \cdots & \boldsymbol{H}_d & \mathbf{0} & \cdots & \mathbf{0} \\ \mathbf{0} & \boldsymbol{H}_0 & \cdots & \boldsymbol{H}_d & \mathbf{0} & \cdots \\ \vdots & & \ddots & & \ddots & \\ \mathbf{0} & \cdots & \mathbf{0} & \boldsymbol{H}_0 & \cdots & \boldsymbol{H}_d \end{bmatrix} \begin{bmatrix} \mathbf{I} \\ \mathrm{q}\mathbf{I} \\ \vdots \\ \mathrm{q}^{k+d}\mathbf{I} \end{bmatrix} = \mathbf{I}.$$

Hence, one can obtain $\boldsymbol{H}^\dagger(\mathrm{q})$ with standard linear algebra by solving

$$\mathrm{blkrow}(\boldsymbol{H}^\dagger(\mathrm{q}))\bar{\boldsymbol{H}} = \begin{bmatrix} \mathbf{I} & \mathbf{0} & \cdots & \mathbf{0} \end{bmatrix}$$

where the number of rows of $\bar{\boldsymbol{H}}$ is $k + 1$.

Let us analyze equation (5) further. By dividing it by $d(\mathrm{q})$, we can distinguish $r$ as the sum of the outputs of four LTI systems. The first term performs a linear combination, where the weights are the faults $f_i$, of the LTI systems $\boldsymbol{N}(\mathrm{q})\boldsymbol{G}_i(\mathrm{q})$ applied to the available signals $\boldsymbol{z}$. The second term contains responses of LTI systems to both $\boldsymbol{\xi}$ and $\boldsymbol{z}$, but scaled to second-order combinations of the faults; hence, they tend to have a negligible effect on the residual if the faults are small. The third and fourth terms show the effect of noise, which, when the faults are small, is dominated by the fourth term $-\boldsymbol{N}(\mathrm{q})\boldsymbol{W}(\mathrm{q})$. These observations indicate that, for any $\boldsymbol{N}(\mathrm{q})$ satisfying the conditions in Lemma 3.1, the regressor generator $d^{-1}(\mathrm{q})\boldsymbol{M}(\mathrm{q})$ where

$$\boldsymbol{M}(\mathrm{q}) := \begin{bmatrix} \boldsymbol{N}(\mathrm{q})\boldsymbol{G}_1(\mathrm{q}) \\ \boldsymbol{N}(\mathrm{q})\boldsymbol{G}_2(\mathrm{q}) \\ \vdots \\ \boldsymbol{N}(\mathrm{q})\boldsymbol{G}_m(\mathrm{q}) \end{bmatrix}, \tag{6}$$

satisfies requirement 2 of Definition 3.2 by taking $\boldsymbol{T}(\mathrm{q}) = -\boldsymbol{N}(\mathrm{q})\boldsymbol{W}(\mathrm{q})$. To additionally satisfy requirement (b), the residual must satisfy the following equivalent linear-algebraic conditions:

**Theorem 3.1 (Linear-algebraic filter characterization).** *Consider system* (1)*, matrices* $\boldsymbol{G}_i$ *as in Lemma 3.1, and assume Assumption 2.1 holds. Given a polynomial* $d(\mathrm{q})$ *of sufficiently high order, the system* $d^{-1}(\mathrm{q})\boldsymbol{N}(\mathrm{q})\boldsymbol{L}(\mathrm{q})$ *is an independently sensitive residual generator with corresponding regressor generator* $d^{-1}(\mathrm{q})\boldsymbol{M}(\mathrm{q})$ *if* $\boldsymbol{M}(\mathrm{q})$ *has the form* (6) *and*

$$\mathrm{blkrow}(\boldsymbol{N}(\mathrm{q}))\bar{\boldsymbol{H}} = \boldsymbol{0}, \tag{7a}$$

$$\mathrm{rank}\,(\mathrm{blkrow}(\boldsymbol{M}(\mathrm{q}))) = m, \tag{7b}$$

*where*

$$\mathrm{blkrow}(\boldsymbol{M}(\mathrm{q})) = \begin{bmatrix} \mathrm{blkrow}(\boldsymbol{N}(\mathrm{q}))\bar{\boldsymbol{G}}_1 \\ \vdots \\ \mathrm{blkrow}(\boldsymbol{N}(\mathrm{q}))\bar{\boldsymbol{G}}_m \end{bmatrix}. \tag{8}$$

*Proof:* See Section 7.1. ■

Unlike residual generator design for additive faults as in [27], conditions (7) do not allow for an efficient linear-programming characterization, by virtue of the rank condition (7b). This rank condition does not appear in [28], since it handles a single multiplicative fault: when $m = 1$, condition (7b) simplifies to $\mathrm{blkrow}(\boldsymbol{N}(\mathrm{q}))\bar{\boldsymbol{G}}_1 \neq \boldsymbol{0}$, which is the same condition as in the design of a filter for additive faults. In addition, in [28], the concern is precisely when the residual generator for a given additive fault *cannot* nullify the effect of the multiplicative fault; that is, the sensitivity to a multiplicative fault is a by-product of the residual design rather than a requirement.

Nonetheless, the matrix $\mathrm{blkrow}(\boldsymbol{M}(\mathrm{q}))$ has $m$ rows and $(k_m + 1)n_{\mathrm{z}}$ columns, where $k_m$ is the degree of $\boldsymbol{M}(\mathrm{q})$; as such, if $(k_m + 1)n_{\mathrm{z}} > m$, it is unlikely that $\boldsymbol{N}(\mathrm{q})$ and $\boldsymbol{M}(\mathrm{q})$ violate (7). Therefore, we suggest the following simple procedure to obtain matrices $\boldsymbol{N}(\mathrm{q})$ and $\boldsymbol{M}(\mathrm{q})$, which works satisfactorily for small $m$:

1. Compute a unitary matrix representation $\boldsymbol{N}_H$ of the left null-space of $\bar{\boldsymbol{H}}$ by, e.g., singular-value decomposition; let $b$ be the number of rows of $\boldsymbol{N}_H$.
2. Randomly generate $K$ unitary vectors $\boldsymbol{v}_i \in \mathbb{R}^b$.
3. For each $i = 1, ..., K$, compute $\mathrm{blkrow}(\boldsymbol{N}^i(\mathrm{q})) = \boldsymbol{v}_i^{\mathsf{T}}\boldsymbol{N}_H$ and $\mathrm{blkrow}(\boldsymbol{M}^i(\mathrm{q}))$ following (8).
4. Compute $i^* = \arg\max_{i \in \mathbb{N}_{\leq K}}(s_{\min}(\mathrm{blkrow}(\boldsymbol{M}^i(\mathrm{q}))))$.
5. Select the filter parameters $\boldsymbol{N}(\mathrm{q}) = \boldsymbol{N}^{i^*}(\mathrm{q})$ and $\boldsymbol{M}(\mathrm{q}) = \boldsymbol{M}^{i^*}(\mathrm{q})$.

The procedure above is a Monte-Carlo optimization where the minimum singular value of $\mathrm{blkrow}(\boldsymbol{M}(\mathrm{q}))$ is maximized. This not only ensures full rank of $\boldsymbol{M}(\mathrm{q})$ but additionally controls its condition number. This concludes the residual and regressor design results.

## 3.2 Fault estimation

With the generators designed according to Theorem 3.1, the main objective of this subsection is to show that our proposed least-squares approach (the rightmost block of Fig. 1) solves Problem 2.1.

Let us call $\boldsymbol{e} := d(\mathfrak{q})^{-1}\boldsymbol{M}(\mathfrak{q})\boldsymbol{z}$ the regressor signal. Throughout this Section, we fix a sampling interval $h$ and a sample count $N$ to be used in the regression step.[1] The fault estimation is performed by computing, at every time step $t \geq Nh$ multiple of $h$, a standard least-squares operation, i.e.,

$$\hat{\boldsymbol{f}}(t) = (\boldsymbol{E}^h_{[t-(N-1)h,t]})^\dagger \boldsymbol{R}^h_{[t-(N-1)h,t]}. \tag{9}$$

where $\boldsymbol{E}^\dagger = (\boldsymbol{E}^\mathsf{T}\boldsymbol{E})^{-1}\boldsymbol{E}^\mathsf{T}$ is the (Moore–Penrose) pseudo-inverse of $\boldsymbol{E}$.[2]

In what follows, we provide performance bounds in the estimation error $\boldsymbol{f} - \hat{\boldsymbol{f}}(t)$. We start by representing (5) in discrete time according to the sampling interval $h$ selected for the fault estimation block:

$$\begin{aligned} r &= \boldsymbol{f}^\mathsf{T}\boldsymbol{e} + \sum_{i,j} f_i f_j \mathscr{T}^J_{i,j}(\mathfrak{z}) \begin{bmatrix} \boldsymbol{\xi} \\ \boldsymbol{z} \end{bmatrix} + \left( \sum_i f_i \mathscr{T}^F_i(\mathfrak{z}) + \mathscr{T}^W(\mathfrak{z}) \right) \boldsymbol{w}, \\ e_i &= \mathscr{T}^M_i(\mathfrak{z})\boldsymbol{z}, \quad i = 1, \ldots, m \end{aligned} \tag{10}$$

where the transfer functions $\mathscr{T}^J_{i,j}(\mathfrak{z}), \mathscr{T}^F_i(\mathfrak{z}), \mathscr{T}^W(\mathfrak{z})$, and $\mathscr{T}^M_i(\mathfrak{z})$ are the exact discrete-time versions (with sampling interval $h$) of $d^{-1}(\mathfrak{q})\boldsymbol{J}_{i,j}(\mathfrak{q}), d^{-1}(\mathfrak{q})\boldsymbol{F}_i(\mathfrak{q})$, $-d^{-1}(\mathfrak{q})\boldsymbol{N}(\mathfrak{q})\boldsymbol{W}(\mathfrak{q})$, and $d^{-1}(\mathfrak{q})\boldsymbol{N}(\mathfrak{q})\boldsymbol{G}_i(\mathfrak{q})$, respectively. To prevent clutter, hereafter we abuse the notation by preserving the notation of signals despite the sampling process.

To deal with noise, we follow the stochastic perturbation theory of [30], using a small-noise approach; i.e., we assume the variances of $\boldsymbol{E}$ and $\boldsymbol{R}$ are, in an appropriate sense, small in comparison to their respective expected values. This enables approximate computations of the expected error norm of an LS estimator with errors in variables based on a first order approximation of the LS estimate, which is asymptotically exact as variances tend to zero (see [30, Theorem 2.8]). Hence, let us make the contribution of noise explicit by denoting $\tilde{\boldsymbol{E}} := \boldsymbol{E} + \boldsymbol{E}_w$ and $\tilde{\boldsymbol{R}} := \boldsymbol{R} + \boldsymbol{R}_w$, where $\boldsymbol{E}, \boldsymbol{R}$ are the noise-free regressor and residual matrices, $\boldsymbol{E}_w, \boldsymbol{R}_w$ are the perturbations to these matrices exclusively by noise, and $\tilde{\boldsymbol{E}}, \tilde{\boldsymbol{R}}$ are the respective perturbed matrices. From (10), let $\boldsymbol{R}_{\mathrm{NL}}$ be the component of the regressor coming from the second-order terms. It holds that

$$\boldsymbol{R} = \boldsymbol{E}\boldsymbol{f} + \boldsymbol{R}_{\mathrm{NL}},$$

while the noisy least squares estimation gives

$$\tilde{\boldsymbol{E}}\hat{\boldsymbol{f}} = \tilde{\boldsymbol{R}}.$$

Thus, the following expression relates the fault estimate and the actual fault:

$$\hat{\boldsymbol{f}} = \tilde{\boldsymbol{E}}^\dagger(\boldsymbol{E}\boldsymbol{f} + \boldsymbol{R}_w) + \tilde{\boldsymbol{E}}^\dagger \boldsymbol{R}_{\mathrm{NL}}, \tag{11}$$

The first-order approximation of the estimate $\hat{\boldsymbol{f}}$, denoted $\breve{\boldsymbol{f}}$, is given by [30][3]

$$\breve{\boldsymbol{f}} = \boldsymbol{f} + \boldsymbol{E}^\dagger\boldsymbol{R}_{\mathrm{NL}} - \boldsymbol{E}^\dagger(\boldsymbol{R}_w - \boldsymbol{E}_w\boldsymbol{f}) - \big((\boldsymbol{E}^\mathsf{T}\boldsymbol{E})^{-1}\boldsymbol{E}_w^\mathsf{T}\boldsymbol{P}_\perp - \boldsymbol{E}^\dagger\boldsymbol{E}_w\boldsymbol{E}^\dagger\big)\boldsymbol{R}_{\mathrm{NL}}, \tag{12}$$

where $\boldsymbol{P}_\perp := \mathbf{I} - \boldsymbol{E}\boldsymbol{E}^\dagger$ is the annihilator (projection) matrix.

Finally, a special type of system norm shall be useful:

[1] If the signals are in discrete time, $h$ must be a natural number.

[2] Because our estimation algorithm is performed every time step, a more computationally efficient online fault estimation implementation should use the recursive least-squares algorithm, see, e.g., [32, Sec. 17.6]. For analysis, we use the standard least-squares solution throughout the text.

[3] Eq. (12) combines expressions from Sections 3.1 and 3.4 of [30] which give the first-order approximations of, respectively, the perturbed pseudo-inverse and the ordinary-least-squares (OLS) estimate under zero-mean perturbation. The combination is necessary because of the term $\boldsymbol{R}_{\mathrm{NL}}$, which is nonzero in general.

**Definition 3.3** ($\mathscr{H}_\infty^\mathrm{F}$ **norm).** Let $\mathscr{T}(\mathrm{q})$ be the transfer function of a BIBO stable system. Its $\mathscr{H}_\infty^\mathrm{F}$ norm is defined as $\sup_{\omega\in\mathbb{R}} \|\mathscr{T}(\mathrm{j}\omega)\|_\mathrm{F}$ if it is a continuous-time system, and as $\sup_{\omega\in[0,2\pi]} \|\mathscr{T}(\exp(\mathrm{j}\omega))\|_\mathrm{F}$ if it is discrete-time.

It is easy to see that the $\mathscr{H}_\infty^\mathrm{F}$ metric enjoys the same norm properties as the Frobenius norm, and hence it is itself a norm. It is also submultiplicative thanks to the submultiplicativity of the Frobenius norm. It can be computed by taking the square root of the $\mathscr{H}_\infty$ norm of the system $\sum_{i,j} \mathscr{T}_{ij}^* \mathscr{T}_{ij} = \mathrm{tr}(\mathscr{T}^*\mathscr{T})$.

Let $\eta_F$ and $\eta_W$ be the $\mathscr{H}_\infty^\mathrm{F}$ norms of the transfer functions $\mathscr{T}^F$ and $\mathscr{T}^W$, respectively, and denote $\eta^2 := \eta_F^2 + \eta_W^2$. In addition, let $\|\mathscr{T}\|_{\infty,\infty}$ be the peak-to-peak gain of $\mathscr{T}$, and denote by $A := \max_{i,j} \|\mathscr{T}_{i,j}^J\|_{\infty,\infty}$. Our main performance result follows.

**Theorem 3.2 (Performance characterization: bias and variance).** *Consider system* (1) *satisfying Assumption 2.1 and further assume it is BIBO stable. Let residual $r$ and regressors $\boldsymbol{e}$ be generated by BIBO stable filters $d^{-1}(\mathrm{q})\boldsymbol{N}(\mathrm{q})\boldsymbol{L}(\mathrm{q})\boldsymbol{z}$ and $d^{-1}(\mathrm{q})\boldsymbol{M}(\mathrm{q})\boldsymbol{z}$, respectively. Furthermore, let $s_i$ be the $i$-th largest singular value of $\boldsymbol{E}$. Then, for any $N$, any fault $\boldsymbol{f}$ and any time $t \geq Nh$, the first-order approximation of the fault estimate $\check{\boldsymbol{f}}$ provided by* (12) *satisfies*

*Bias:*
$$\left|\mathbb{E}(\check{\boldsymbol{f}}(t) - \boldsymbol{f})\right| \leq A\sqrt{N} m |\boldsymbol{f}|^2 s_m^{-1} \left\|\begin{bmatrix}\boldsymbol{\xi}^\mathsf{T} & \boldsymbol{z}^\mathsf{T}\end{bmatrix}\right\|_\infty =: B, \tag{13a}$$

*Variance:*
$$\mathrm{tr}(\mathrm{Var}(\check{\boldsymbol{f}}(t) - \boldsymbol{f})) \leq \sigma^2 \left( 2(|\boldsymbol{f}|^2 + 1)\eta^2 \sum_{i=1}^m s_i^{-2} + B^2 \eta_F^2 \left( 2\sum_{i=1}^m s_i^{-2} + s_m^{-2} \right)\right). \tag{13b}$$

*Proof:* See Section 7.2. ■

Since $A$ is the maximal peak-to-peak norm among $\mathscr{T}_{i,j}^J(\mathrm{q}) = d(\mathrm{q})^{-1}\boldsymbol{J}_{i,j}(\mathrm{q})$, Theorem 3.2 with Lemma 3.1 give the following corollary:

**Corollary 3.1 (Bias and variance when faults multiply only known signals).** *Given the premises of Theorem 3.2, suppose additionally that $\boldsymbol{H}_i'(\mathrm{q}) \equiv \boldsymbol{0}$ for all $i = 1, \dots, m$. Then, more simply, $\left|\mathbb{E}(\check{\boldsymbol{f}}(t) - \boldsymbol{f})\right| = 0$ and $\mathbb{E}(|\check{\boldsymbol{f}}(t) - \boldsymbol{f}|^2) \leq (|\boldsymbol{f}|^2 + 1)\eta^2\sigma^2 \sum_{i=1}^m s_i^{-2}$.*

The involved expressions (13a) and (13b) require some unpacking. We start by the simpler case of Corollary 3.1, where no errors due to nonlinearities occur. Compare it with ordinary least squares (OLS), where the linear model $\boldsymbol{R} = \boldsymbol{E}\boldsymbol{f} + \boldsymbol{v}$, with i.i.d. $v_i$ with zero mean and variance $\sigma^2$, gives an error variance asymptotically equal to $\sigma^2(\boldsymbol{E}^\mathsf{T}\boldsymbol{E})^{-1}$. Hence, its trace, which is the expected value of the squared Euclidean norm of the error, converges to $\sigma^2 \sum_{i=1}^m s_i^{-2}$. Our result is the same, but amplified by the factor $(1 + |\boldsymbol{f}|^2)$ and further scaled by the squared $\mathscr{H}_\infty^\mathrm{F}$ norm of the systems that filter the white noise $\boldsymbol{w}$. The $\mathscr{H}_\infty^\mathrm{F}$-norm essentially measures the amplification/attenuation of noise that the systems provide, but also accounts for the added autocorrelation generated by filtering white noise. The amplification factor $(1 + |\boldsymbol{f}|^2)$ is due to the errors-in-variables nature, and is consistent with the observations of, e.g., [30, Sec. 3.4].

Now, let us inspect the bias term of (13a). As expected from the first-order approach to the estimation problem, it increases with the square of fault magnitude, scaled by the peak-to-peak norm of associated systems $d(\mathrm{q})^{-1}\boldsymbol{J}_{i,j}(\mathrm{q})$ which map the unmeasured variables into the residual, and the magnitude of the signals involved. It also increases with the number of faults $m$. However, it *decreases* with $s_m/\sqrt{N}$, which can be seen as an effective regressor richness measurement, see Remark 3.3.

Returning to the explanation of the total variance bound, (13b) includes a second term, which is the squared bias given in (13a) multiplied by the factor $\eta_F^2(2\sum_{i=1}^m s_i^{-2} + s_m^{-2}) \leq \eta_F^2(2m+1)s_m^{-2}$. Three observations are in order: (i) this term is the most negligible when faults are small, given the factor $|\boldsymbol{f}|^4$; (ii) this term is essentially a multiple of $s_m^{-4}$, while the previously discussed term is a multiple only of $s_m^{-2}$; thus, a large value of $s_m$ should also contribute to the low importance of this term; and (iii) this term does increase linearly with $N$, which is not the case for the previous term.

**Remark 3.2 (Relation to Gauss–Newton method for nonlinear least squares).** The problem of finding $f_i$ given $\boldsymbol{z}$ using (5) can be seen as a nonlinear least-squares (NLS) problem, perturbed by $\boldsymbol{\xi}$. Therefore, using the linear-least-squares approximation in (9) is similar to performing one step of the Gauss–Newton method for the NLS problem involved. For this reason, a natural extension of the fault estimation approach proposed here is to iteratively (i) compute residual and regressor generators using Theorem 3.1; (ii) run the fault estimation algorithm using Eq. (9) up to a sufficiently long time $T$; update the nominal polynomial matrices using $[\boldsymbol{H}(\mathfrak{q}) \quad \boldsymbol{L}(\mathfrak{q})] \leftarrow [\boldsymbol{H}(\mathfrak{q}) \quad \boldsymbol{L}(\mathfrak{q})] + \sum_i \hat{f}_i(T)[\boldsymbol{H}_i'(\mathfrak{q}) \quad \boldsymbol{L}_i'(\mathfrak{q})]$. Under similar convergence conditions as the nominal Gauss–Newton method for NLS (see, e.g., [32, Section 17.4]), the successive incremental fault estimates and residuals converge to zero, and the total fault is the cumulative sum of the incremental estimates. It is important to highlight that, while this process involves switching the dynamics of the generators in real time, it does not hinder their stability, as the poles, dictated by $d(\mathfrak{q})$, are unchanged.[4]

**Remark 3.3. (Effective singular values)** Observation (iii) above may give the impression that, contrary to intuition, more data can actually have a negative impact in the estimation performance, but this is not correct. Generally, $s_{\min}(\boldsymbol{E})$ grows linearly with $\sqrt{N}$. E.g., it is easy to verify that the singular values of $\begin{bmatrix}\boldsymbol{E} & \boldsymbol{E} & \cdots \boldsymbol{E}\end{bmatrix}$, where $\boldsymbol{E}$ is repeated $k$ times, are $\sqrt{k}$ times the singular values of $\boldsymbol{E}$. Hence, when $\boldsymbol{e}(t)$ is periodic, $\lim_{N\to\infty} \mathbb{E}(|\breve{\boldsymbol{f}}(t) - \boldsymbol{f}|^2) = 0$, provided $\boldsymbol{E}$ is full-rank. This asymptotic result does not hold for the bias in (13a): longer experiments cannot cancel the bias introduced by a linear model not properly representing the relation between residual and regressors, which is a known fact for nonlinear regression. Therefore, one can call $s_m/\sqrt{N}$ an effective singular value that measures the richness of data irrespective of its size.

**Remark 3.4. (Signal-to-noise ratio)** Theorem 3.2 is an asymptotic result using a small-noise assumption. In [30], a metric is suggested to assess whether the first-order approximation for least squares is accurate. In our case, it has the following upper bound (see Section 7.3 for the derivation):

$$c \leq \sqrt{2N} s_m^{-1} \sigma \gamma_F. \tag{14}$$

Due to the asymptotic arguments that justify the theory in [30], this metric must be significantly smaller than 1 for Theorem 3.2 to have practical meaning. Thus, the inverse $c^{-1}$ can be interpreted as a signal-to-noise ratio (SNR), where the signal is measured by the effective richness $s_m/\sqrt{N}$ and the noise by $\sigma\gamma_F$.

Theorem 3.2 indicates the main drivers of estimation performance, among which the smallest singular value of $\boldsymbol{E}$ is the only one depending on real-time deterministic signals. This is the starting point for the experiment design problem we present in Section 4.

## 3.3 The case of time-varying faults

The approach outlined in this section has only been analyzed for constant faults. However, it can be applied to time-varying faults of the form $f_i = \sum_j p_{ij}\phi_{ij}(t)$, where $p_{ij}$ are unknown parameters and $\phi_{ij}(t)$ are known signals. Its application needs a slight modification, which we explain next.

To accommodate the use of time signals and the $\mathfrak{q}$ operator, we introduce a notational convention. Throughout this text, the pre-multiplication of a signal with another signal shall be understood as an operator. That is, $a(t)b(t)$ is the result of applying the operator that multiplies signal $b$ by signal $a$. The main observation is that multiplication is commutative but differentiation does not commute with multiplication. That is, in continuous time, for signals $a$ and $b$, $ab = ba$ but $a\mathfrak{q}b \neq \mathfrak{q}ab$. Hence, we shall always read $abc := a(b(c))$ as usual when applying operators, where $a, b, c$ might be either signals to be multiplied or (differentiation/shift) operators.

With this operator approach, the following identities are important:

(i) *continuous-time:* for signals $a$ and $b$, because $\mathfrak{q}(ab) = (\mathfrak{q}a)b + a\mathfrak{q}b$, we write

$$\mathfrak{q}a = \dot{a} + a\mathfrak{q}. \tag{15}$$

(ii) *discrete-time:* for sequences $a$ and $b$, because $\mathfrak{q}(ab) = (\mathfrak{q}a)(\mathfrak{q}b)$, we write

$$\mathfrak{q}a = a(k+1)\mathfrak{q} \tag{16}$$

[4]Stability of switched linear systems can be assessed by finding a common Lyapunov function for all systems. If all modes have the same poles, they admit a state-space realization with the same $\boldsymbol{A}$ matrix; since the Lyapunov function depends only on $\boldsymbol{A}$, stability is retained with switching zeros.

Modulo the order of operations in the second order terms, one can see that Lemma 3.1 still holds in the time-varying fault case; the only difference is that the second-order term in (5) becomes

$$N(\mathfrak{q})\sum_{i,j} f_i(t)H_i'(\mathfrak{q})H^\dagger(\mathfrak{q})f_j(t)H_j'(\mathfrak{q}).$$

Therefore, the linear approximation still holds, and, as such, the estimation performance has the same essential characteristics as the constant-fault presented above.

The modification is required on the implementation side. It must remain possible to generate the regressor signals from measured signals and signals that are known in advance. In the time-varying case, the known signals are $z$, and, since $\phi_{ij}$ is known a priori, potentially higher-order derivatives of $\phi_{ij}$ can be also used. Since the regressors come from the linear term in (5):

$$d^{-1}(\mathfrak{q})N(\mathfrak{q})\sum_i f_i(t)G_i(\mathfrak{q}) = \sum_{ij} p_{ij} d^{-1}(\mathfrak{q})N(\mathfrak{q})\phi_{ij}G_i(\mathfrak{q}),$$

we must be able to implement the regressors

$$d^{-1}(\mathfrak{q})N(q)\phi_{ij}G_i(q).$$

To obtain a realizable filter that avoids derivatives of the measurements, we can simply apply the operations (i) or (ii) above repeatedly so that the regressor ends with the form $d^{-1}(\mathfrak{q})N(\mathfrak{q})G_{ij}'(\mathfrak{q})\phi_{ij}'$, where $\phi'$ may contain higher-order derivatives or time advances. When building the regressor, one needs to perform multiplications of $z$ with $\phi_{ij}'$ accordingly, and then feed each of these signals to the corresponding LTI filter $d^{-1}N(\mathfrak{q})G_{ij}'(\mathfrak{q})$.

This approach is better understood with an example.

**Example 3.1.** Consider a single fault of the form $f(t) = p_1 + p_2 t$, and let

$$G(\mathfrak{q}) = \begin{bmatrix} \mathfrak{q}^2 + \mathfrak{q} + 1 & 1 \end{bmatrix}.$$

The only non-trivial element of $G(\mathfrak{q})$ is the first. The two basis functions are $\phi_1(t) = 1$ (constant) and $\phi_2(t) = t$. Applying identity (i), we get $t\mathfrak{q} = \mathfrak{q}t - 1$ and $t\mathfrak{q}^2 = (\mathfrak{q}t - 1)\mathfrak{q} = \mathfrak{q}(t\mathfrak{q}) - \mathfrak{q} = \mathfrak{q}(\mathfrak{q}t - 1) - \mathfrak{q} = \mathfrak{q}^2 t - 2\mathfrak{q}$. Therefore, $t(\mathfrak{q}^2 + \mathfrak{q} + 1) = \mathfrak{q}^2 t - 2\mathfrak{q} + \mathfrak{q}t - 1 + t = (\mathfrak{q}^2 + \mathfrak{q} + 1)t - 2\mathfrak{q} - 1$. Hence, $\phi_1'(t) = 1$, $\phi_2'(t) = t$, and

$$G_{11}'(q) = \begin{bmatrix} \mathfrak{q}^2 + \mathfrak{q} + 1 & 1 \end{bmatrix},\ G_{12}'(q) = \mathbf{0},\ G_{21}'(q) = \begin{bmatrix} -2\mathfrak{q} - 1 & 0 \end{bmatrix},\ G_{22}'(q) = \begin{bmatrix} \mathfrak{q}^2 + \mathfrak{q} + 1 & 1 \end{bmatrix}.$$

Given these matrices, the regressor for $p_1$ is generated by $d^{-1}(\mathfrak{q})N(\mathfrak{q})G_{11}'(\mathfrak{q})z$, while the regressor for $p_2$ is generated by $d^{-1}(\mathfrak{q})N(\mathfrak{q})G_{21}'(\mathfrak{q})z + d^{-1}(\mathfrak{q})N(\mathfrak{q})G_{22}'(\mathfrak{q})tz$.

# 4 Input design for active estimation

In this section, we describe an optimization problem that gives the input design that maximizes the fault estimator performance, followed by our solution approach.

## 4.1 Optimization problem

Theorem 3.2 tells that the performance of the fault estimator is inversely proportional to $s_{\min}(E)$. Thus, it makes sense to design an input $u$ that maximizes the minimum singular value of $E$, a metric we shall denote as *richness*. Unfortunately, the regressor signal $e$ is affected by all signals entering the system, including the potential faults $f$. Nevertheless, we chose to design the input assuming the nominal case, i.e., $f = 0, w \equiv \mathbf{0}, d \equiv \mathbf{0}$, where $d$ is the external disturbance within the partition of $\xi$. Under this assumption, $e$ is a function of $u$.

Recall that the signals for the estimation are sampled with period $h$. We shall hereafter assume that an exact discrete-time version $\mathscr{T}_Y(\mathfrak{z})$ of the nominal system of (1) is available, i.e., $y = \mathscr{T}_Y(\mathfrak{z})u$. Then, by letting $\mathscr{T}_M(\mathfrak{z})$ be the exact discrete-time representation of $d^{-1}(q)M(\mathfrak{q})$, we have

$$e = \mathscr{T}(\mathfrak{z})u := \mathscr{T}_M(\mathfrak{z})\begin{bmatrix} \mathscr{T}_Y(\mathfrak{z}) \\ \mathbf{I} \end{bmatrix} u. \tag{17}$$

Before writing down the optimization problem, we must impose a few design choices. First, we remark that a finite-horizon experiment design problem is ill-advised, unless the initial state of the system can be ensured to be zero. This is due to the uncontrollable effect of the unknown initial state in the richness of $\boldsymbol{e}$. Instead, because the system and filter are stable, it is more convenient to assume that the experiment runs for sufficient long time so that the contribution of the initial state vanishes. Therefore we suggest a moving-horizon approach: we want that $\boldsymbol{U}^h_{[t-(N-1)h,t]}$ maximizes the richness $s_{\min}(\boldsymbol{E}^h_{[t-(N-1)h,t]})$ for sufficiently large $t$. This asymptotic richness exists if $\boldsymbol{e}$ is asymptotically periodic, as we show next:

**Proposition 4.1 (Asymptotic richness for periodic inputs).** *If $\boldsymbol{e}$ is asymptotically $N$-periodic, then the limit $\lim_{t\to\infty} s_{\min}(\boldsymbol{E}_{[t,t+N-1]})$ exists.*

*Proof:* See Section 7.5. ∎

Because $\mathscr{T}(\mathfrak{z})$ is a linear operator, one could always increase the magnitude of $\boldsymbol{U}$ to increase $s_{\min}(\boldsymbol{E})$. To make the problem well-posed (and realistic), we assume that $\boldsymbol{U}$ has to be constrained to some compact and bounded set $\mathscr{U}$. We shall further assume that $\mathscr{U}$ is convex. For simplicity, assume further that $h=1$. We can define the optimal input design problem as follows:

**Definition 4.1 (Optimal input for asymptotic richness).** Consider (17) and associated system (1). Assume $\boldsymbol{u}$ is an $N$-periodic discrete-time signal and denote the *asymptotic richness* by $J_\infty(\boldsymbol{u}) := \lim_{t\to\infty} s_{\min}(\boldsymbol{E}_{[t,t+N-1]})^2$. An $N$-periodic input signal $\boldsymbol{u}^\star$ in a given closed bounded convex subset $\mathscr{U}$ is said to be *optimal for fault estimation* if $\boldsymbol{u}^\star = \arg\max_{\boldsymbol{u}\in\mathscr{U}} J_\infty(\boldsymbol{u})$.

Hereafter, let us use the shorthand $\boldsymbol{U} := \boldsymbol{U}_{[0,N-1]}$, and let $\bar{\boldsymbol{u}} \in \mathbb{R}^{Nn_{\mathrm{u}}}$ be the column vector obtained by stacking the columns of $\boldsymbol{U}$. Furthermore, let $\bar{\mathscr{U}}$ be its associated convex constraint set; since $\boldsymbol{u}$ is $N$-periodic by definition, there exists a natural bijection between $\bar{\mathscr{U}}$ and $\mathscr{U}$ of Def. 4.1. Consider now a minimal state-space realization of $\mathscr{T}(\mathfrak{z})$ and its matrices $\boldsymbol{A},\boldsymbol{B},\boldsymbol{C},\boldsymbol{D}$, and define the following matrices for all $i=1,...,N$:

$$\begin{aligned}
\boldsymbol{P}_i &:= \begin{bmatrix} \boldsymbol{C}\boldsymbol{A}^{i-2}\boldsymbol{B} & \cdots & \boldsymbol{C}\boldsymbol{B} & \boldsymbol{D} & 0_{m\times n_{\mathrm{u}}(N-i)} \end{bmatrix},\\
\boldsymbol{P}_x &:= \begin{bmatrix} \boldsymbol{A}^{N-1}\boldsymbol{B} & \boldsymbol{A}^{N-2}\boldsymbol{B} & \cdots & \boldsymbol{A}\boldsymbol{B} & \boldsymbol{B} \end{bmatrix},\\
\boldsymbol{P}'_i &:= \boldsymbol{P}_i + \boldsymbol{C}\boldsymbol{A}^{i-1}(\mathbf{I}-\boldsymbol{A}^N)^{-1}\boldsymbol{P}_x.
\end{aligned} \tag{18}$$

Our main optimization result follows:

**Theorem 4.1 (Objective function and first-order information).** *Consider* (17)*,* (18)*, and associated system* (1)*, and assume $\mathscr{T}(\mathfrak{z})$ is BIBO stable. Let $\lambda_{\min}, \boldsymbol{v}_{\min}$ be a smallest eigenvalue and associated unitary eigenvector of the p.s.d. matrix $\mathbf{Q}(\bar{\boldsymbol{u}}) := \sum_{i=1}^N \boldsymbol{P}'_i\bar{\boldsymbol{u}}\bar{\boldsymbol{u}}^\mathsf{T}\boldsymbol{P}'^\mathsf{T}_i$, and define the expressions*

*Objective function:*
$$J(\bar{\boldsymbol{u}}) := \lambda_{\min}(\mathbf{Q}(\bar{\boldsymbol{u}})), \tag{19a}$$

*Subgradient:*
$$\boldsymbol{g}(\bar{\boldsymbol{u}}) := 2\sum_{i=1}^N \boldsymbol{v}^\mathsf{T}_{\min}\boldsymbol{P}'_i\bar{\boldsymbol{u}}\boldsymbol{P}'^\mathsf{T}_i\boldsymbol{v}_{\min}. \tag{19a}$$

*Let $\boldsymbol{u}$ be the signal built by the periodic repetition of the elements in $\bar{\boldsymbol{u}}$, i.e., $u_i(k) = \bar{u}_{i+n_{\mathrm{u}}(k \bmod N)}$ for all $i = 1,..,n_{\mathrm{u}}, k\in\mathbb{N}$; here,* mod *denotes the remainder operator. Then, $J(\bar{\boldsymbol{u}}) = J_\infty(\boldsymbol{u})$ (Def. 4.1) and $\boldsymbol{g}(\bar{\boldsymbol{u}})$ is a subgradient of $J(\bar{\boldsymbol{u}})$.*

*Proof:* See Section 7.5. ∎

Theorem 4.1 enables us to translate the optimal input design of Definition 4.1 into a finite-dimensional optimization problem. Moreover, it provides a subgradient, which is needed as the minimal eigenvalue function is not everywhere differentiable. With the first-order information, convex optimization methods such as projected subgradient ascent can be used.

**Remark 4.1 (Convex constraints).** Common choices for $\mathscr{U}$ are those described by the following inequalities:

- Component bounds, i.e., $u^L_i \le u_{i,t} \le u^U_i, \forall t\in\{1,2,...,N\}, i\in\{1,2,...,m\}$;
- Component energy bounds, i.e., $\boldsymbol{u}^\mathsf{T}_i\boldsymbol{u}_i \le E_i, \forall i\in\{1,2,...,m\}$;

- Total energy bounds, i.e., $\sum_{i=1}^{m} \boldsymbol{u}_i^\top \boldsymbol{u}_i \leq E$.

Nominal output constraints are also possible, but they may need to be robustified against unknown initial state and disturbance.

**Remark 4.2.** Because the function $f(\boldsymbol{E}) := \lambda_{\min}(\boldsymbol{E}^\top \boldsymbol{E})$ satisfies $f(a\boldsymbol{E}) = a^2 f(\boldsymbol{E})$, the solution to Problem (19) is always on the boundary of $\mathcal{U}$, provided $\mathcal{U}$ is star-shaped at the origin. This is the case, for example, of non-empty convex sets containing the origin. By continuity of $f$, this implies that it has at least one maximum and one saddle point at the boundary, unless $f$ is constant at the boundary of $\mathcal{U}$.

## 4.2 Optimization solution

We propose Algorithm 1, a projected subgradient ascent algorithm for an efficient local-optimal solution to the input design problem. In this method, at each iteration a subgradient of the objective function is computed, a step is taken in such direction, the solution is projected onto the feasible set, and the iteration repeats.

**Algorithm 1** Computation of optimal input for (19)

**Input:** $\boldsymbol{P}_i, i = 1..N, \bar{\mathcal{U}}, \tau, L$
**Output:** $\bar{\boldsymbol{u}}$

1: $k \leftarrow 0$
2: $\bar{\boldsymbol{u}} \leftarrow \text{project}_{\bar{\mathcal{U}}}(\text{rand}(N n_u))$
3: **while** true **do**
4: $\quad \boldsymbol{M} \leftarrow \sum_{i=1}^{N} \boldsymbol{P}_i \bar{\boldsymbol{u}} \bar{\boldsymbol{u}}^\top \boldsymbol{P}_i^\top$
5: $\quad (\lambda_{\min}, \boldsymbol{v}_{\min}) \leftarrow \text{eig}_{\min}(\boldsymbol{M})$
6: $\quad \boldsymbol{g} \leftarrow \sum_{i=1}^{N} \boldsymbol{v}_{\min}^\top \boldsymbol{P}_i \bar{\boldsymbol{u}} \boldsymbol{P}_i^\top \boldsymbol{v}_{\min}$
7: $\quad \bar{\boldsymbol{u}}_{\text{new}} \leftarrow \text{project}_{\bar{\mathcal{U}}}(\bar{\boldsymbol{u}} + \frac{L}{L+k} \tau \boldsymbol{g}^\top)$
8: $\quad$ **if** $|\bar{\boldsymbol{u}}_{\text{new}} - \bar{\boldsymbol{u}}|_2 < \varepsilon_u$ **or** $|\lambda_{\min}|/k < \varepsilon_\lambda$ **then**
9: $\qquad \bar{\boldsymbol{u}} \leftarrow \bar{\boldsymbol{u}}_{\text{new}}$
10: $\qquad$ **return**
11: $\quad$ **end if**
12: $\quad k \leftarrow k + 1$
13: $\quad \bar{\boldsymbol{u}} \leftarrow \bar{\boldsymbol{u}}_{\text{new}}$
14: **end while**

Let us explain Algorithm 1 relying on Theorem 4.1. After $\bar{\boldsymbol{u}}$ is initialized,[5] the main subgradient ascent loop begins. Lines 4–6 use (19a) to compute the subgradient $\boldsymbol{g}$; next, Line 7 computes the projected subgradient update with step size parameters $\tau$ and $L$, using $\text{project}_{\mathcal{U}}$ to perform an orthogonal projection onto the convex set $\mathcal{U}$; finally, Line 8 presents the two stopping criteria: either the change in the Euclidean norm of the gradient below the tolerance $\varepsilon_{\text{u}}$, or the change in the ergodic mean of the objective function $\lambda_{\min}$ is smaller than the tolerance $\varepsilon_\lambda$. Note that, for many practical convex set descriptions, projection is an easy computation; e.g., in the unit ball case, $\bar{\mathcal{U}} = \{\boldsymbol{u} \mid |\boldsymbol{u}|_2 \leq 1\}$, it is computed by normalizing $\boldsymbol{u}$ to unit Euclidean norm when bigger than 1. In addition, we apply a decaying step size of the form $\frac{L}{L+k}\tau$, which has generally improved convergence properties. For the same reason, the ergodic mean is used as stopping criterion.

Since we have presented a local method, we have no guarantee that the global optimum is attained. Thus, it is sensible to estimate an upper bound to the global optimal value, thereby obtaining an over-approximation of the optimality gap. This can be achieved by semidefinite relaxation. Let us assume for simplicity that the constraint set has the form $\bar{\mathcal{U}} = \{\bar{\boldsymbol{u}} \mid \bar{\boldsymbol{u}}^\top \boldsymbol{S}_i \bar{\boldsymbol{u}} \leq 1, i = 1, ..., N_c\}$, where $N_c$ is the number of constraints and $\boldsymbol{S}_i \succeq \boldsymbol{0}$. This is valid for all constraints laid out in Remark 4.1. We can assert the following.

**Proposition 4.2 (Semidefinite relaxation of the optimal experiment design).** *Let $\lambda^\star$ be the value of* (19),

[5] We have chosen to initialize $\bar{\boldsymbol{u}}$ to a pseudorandomly generated vector, but other initalization is possible, as long as it is not the zero vector, which is a global minimum with zero gradient.

*where* $\bar{\mathscr{U}} = \{\bar{\boldsymbol{u}} \mid \bar{\boldsymbol{u}}^{\mathsf{T}}\boldsymbol{S}_i\bar{\boldsymbol{u}} \leq 1, i = 1,...,N_c\}$, *where* $\boldsymbol{S}_i \succeq \boldsymbol{0}$. *Then, the value* $\lambda^{\mathrm{SDP}}$ *of the SDP*

$$\begin{aligned}
\max_{\bar{U},\lambda} \quad & \lambda \\
\text{subject to} \quad & \sum_{i=1}^{N} \boldsymbol{P}_i'\bar{\boldsymbol{U}}\boldsymbol{P}_i'^{\mathsf{T}} \succeq \lambda \boldsymbol{I}, \\
& \operatorname{tr}(\boldsymbol{S}_i\bar{\boldsymbol{U}}) \leq 1, \quad \forall i = 1,...,N_c, \\
& \bar{\boldsymbol{U}} \succeq \boldsymbol{0}
\end{aligned} \tag{20}$$

*is larger than or equal to* $\lambda^\star$. *Furthermore, if* $\operatorname{rank}(\bar{\boldsymbol{U}}) = 1$, *then* $\lambda^{\mathrm{SDP}} = \lambda^\star$.

*Proof:* See Section 7.5 ■

We must remark that the SDP (20) has as its main variable a square $Nn_{\mathrm{u}}$-dimensional matrix. With large values of $N$, it may be challenging to solve it reliably.

# 5 Numerical results

This section presents a series of numerical experiments to illustrate the performance of our method. A Matlab implementation of our methods, including the scripts to reproduce these results, can be found in `https://gitlab.tudelft.nl/ggleizer/aemf`.

## 5.1 Setup

We consider a continuous-time linearized pendulum-cart system, adapted from [33, Eqs. (3) and (6)], with three parametric faults and one external (unmeasured) disturbance described by the following nominal DAE of the form (1):

$$\begin{bmatrix} (M+M_{\mathrm{p}})\mathrm{q}+b & M_{\mathrm{p}}l\mathrm{q}^2 & (M+M_{\mathrm{p}})g \\ M_{\mathrm{p}}l\mathrm{q} & (I+M_{\mathrm{p}}l^2)\mathrm{q}^2+M_{\mathrm{p}}gl & -M_{\mathrm{p}}gl \\ 1 & 0 & 0 \\ 0 & 1 & 0 \end{bmatrix}\boldsymbol{\xi} + \begin{bmatrix} 0 & 0 & -1 & 0 \\ 0 & 0 & 0 & -0.1 \\ -1 & 0 & 0 & 0 \\ 0 & -1 & 0 & 0 \end{bmatrix}\boldsymbol{z} + \begin{bmatrix} \boldsymbol{0} \\ \mathbf{I} \end{bmatrix}\boldsymbol{w} = \boldsymbol{0},$$

where $M$ is the cart mass, $M_{\mathrm{p}}, l, I$ are the nominal pendulum mass, length, and moment of inertia, respectively, $b$ is the nominal friction coefficient, and $g$ is the acceleration of gravity.[6] The linearization is performed at downright pendulum position, hence the system is open-loop stable. The measured signals $\boldsymbol{z}(t) \in \mathbb{R}^4$ can be partitioned as $\begin{bmatrix}\boldsymbol{y}^{\mathsf{T}} & \boldsymbol{u}^{\mathsf{T}}\end{bmatrix}^{\mathsf{T}}$ where $\boldsymbol{y}(t) \in \mathbb{R}^2$ are the car velocity and pendulum angle measurements, and $\boldsymbol{u}(t) \in \mathbb{R}^2$ are the force applied to move the cart horizontally and a torque applied to the base of the pendulum, in the respective order. Likewise, the unmeasured signals $\boldsymbol{\xi}(t) \in \mathbb{R}^4$ can be partitioned as $\begin{bmatrix}\boldsymbol{x}^{\mathsf{T}} & \boldsymbol{d}^{\mathsf{T}}\end{bmatrix}^{\mathsf{T}}$ where $\boldsymbol{x}(t) \in \mathbb{R}^3$ are the internal states (car velocity, pendulum angle, and pendulum angular velocity, in order), and $\boldsymbol{d}(t) \in \mathbb{R}^1$ is the unmeasured slope in which the cart is travelling, being zero when the car moves horizontally.

The fault matrices are

$$\boldsymbol{H}_1'(\mathrm{q}) = \begin{bmatrix} b & 0 & 0 \\ 0 & 0 & 0 \\ 0 & 0 & 0 \\ 0 & 0 & 0 \end{bmatrix}, \quad \boldsymbol{H}_2'(\mathrm{q}) = \begin{bmatrix} M_{\mathrm{p}}\mathrm{q} & M_{\mathrm{p}}l\mathrm{q}^2 & M_{\mathrm{p}} \\ M_{\mathrm{p}}l\mathrm{q} & M_{\mathrm{p}}gl+M_{\mathrm{p}}l^2\mathrm{q}^2 & -M_{\mathrm{p}}g \\ 0 & 0 & 0 \\ 0 & 0 & 0 \end{bmatrix}, \quad \boldsymbol{H}_3'(\mathrm{q}) = \boldsymbol{0},$$

$$\boldsymbol{L}_1'(\mathrm{q}) = \boldsymbol{L}_2'(\mathrm{q}) = \boldsymbol{0}, \quad \boldsymbol{L}_3'(\mathrm{q}) = \begin{bmatrix} 0 & 0 & -1 & 0 \\ 0 & 0 & 0 & -0.1 \\ 0 & 0 & 0 & 0 \\ 0 & 0 & 0 & 0 \end{bmatrix}.$$

[6] See [33] for nominal values and units.

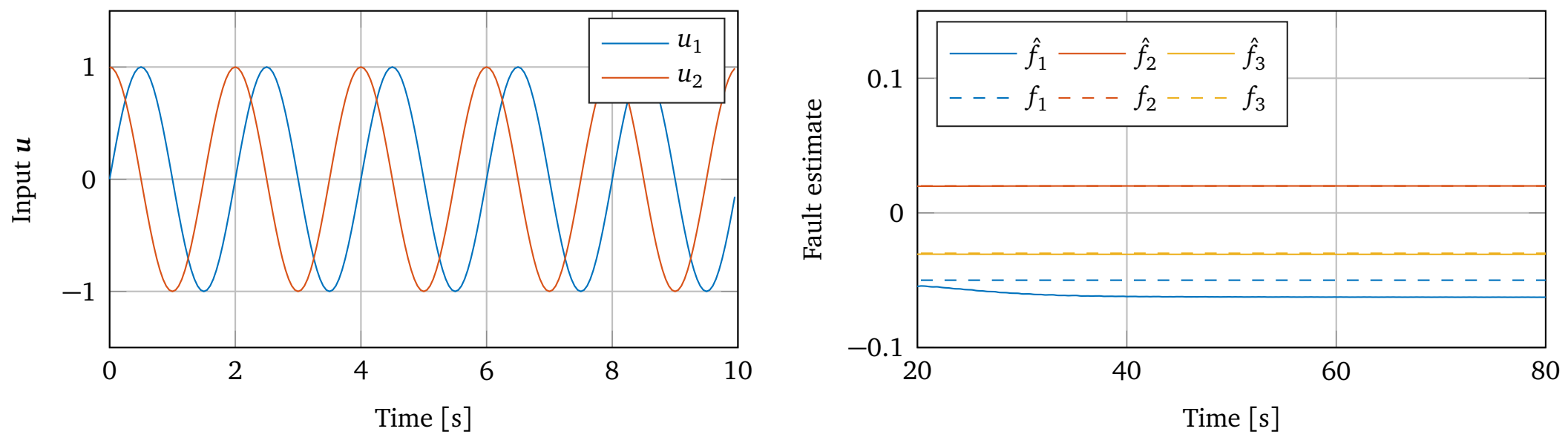

Figure 2: Input (left) and fault estimation results (right) for the first scenario, with simple sinusoidal input and zero noise.

Fault $f_1$ represents a multiplicative change in the friction of the system; fault $f_2$ represents a multiplicative change in the mass of the pendulum; and fault $f_3$ represents a multiplicative factor applied to the battery power of the cart, affecting both force and torque, simultaneously.

## 5.2 Fault estimator and input design

The matrix $\boldsymbol{H}(\mathfrak{q})$ is verified to satisfy Assumption 2.1 with the help of Remark 3.1, where $\boldsymbol{H}^{\dagger}(\mathfrak{q})$ has degree 1. We then solve the equations in Theorem 3.1, which, for degree 2, has the unique solution (up to a scalar)

$$\boldsymbol{N}(\mathfrak{q}) = \begin{bmatrix} -0.065 & 0.026\mathfrak{q} & 0.026 & -0.024 - 0.065\mathfrak{q} & 0.994 + 0.026\mathfrak{q}^2 \end{bmatrix},$$

from which $\boldsymbol{M}(\mathfrak{q})$ can be readily derived following (8), having degree 2. To design the denominator $d(\mathfrak{q})$, whose degree must be bigger than or equal to 2 for a proper realization, we notice that the nominal system has a peak frequency response at 5.58 rad/s, after which it decays rapidly. Therefore, we choose the poles to be higher than this frequency, and set $d(\mathfrak{q}) = (\mathfrak{q}+10)(\mathfrak{q}+20)/200$. For the fault estimator, we have chosen the sampling interval of $h = 0.05$ and estimation window of $N = 400$.

For the input design, we have picked $N = 40$ and chose to bound the 2-norm of each input signal by $\sqrt{N/2}$, so that it has the same 2-norm as a sinusoidal of amplitude 1 and period $N$.[7] We have run Alg. 1 with step size parameters $\tau = 10$ and $L = 50$, and tolerances $\varepsilon = 10^{-3}$ and $\varepsilon_{\lambda} = 10^{-5}$. The algorithm stopped after 3961 iterations, taking only 0.20 second, and yielding an optimal value of 0.212. We used CVX [34] in Matlab to solve the SDP in Prop. 4.2, which took 0.85 second to compute the upper bound to the global optimum of 0.451. This means that the global optimal singular value is between the square roots 0.460 and 0.672.

## 5.3 Results

Now we present the estimation results for multiple simulations, including small constant faults, large constant faults, and time-varying faults.

### 5.3.1 Small faults

We have simulated four scenarios. In all of them, the faults are $f_1 = -0.05, f_2 = 0.02, f_3 = -0.03$, and a disturbance $\boldsymbol{d}(t) = \frac{5\pi}{180}\sin(\pi t)$, corresponding to a terrain oscillating between –5 and 5 degrees of slope. In the first two scenarios we emulate passive fault estimation by setting the input $\boldsymbol{u}(t) = \begin{bmatrix} \sin(\pi t) & \cos(\pi t) \end{bmatrix}^{\mathsf{T}}$, and set noise to zero. In the second scenario, we introduce zero-mean Gaussian white noise with variance 1 through pseudo-randomly generated numbers. In the third and fourth scenarios, we use the optimal input designed using Algorithm 1 for the noiseless case and the noisy output case, respectively.

In the first scenario, without noise, the fault estimation attains a small estimation error, with a more significant bias in the estimation of the friction fault $f_1$, see Fig. 2. The non-zero error is in line with Corollary 3.1, since

[7] It is not necessary that the time window for estimation is the same as that of the input signal, provided the former is an integer multiple of the latter to ensure periodicity.

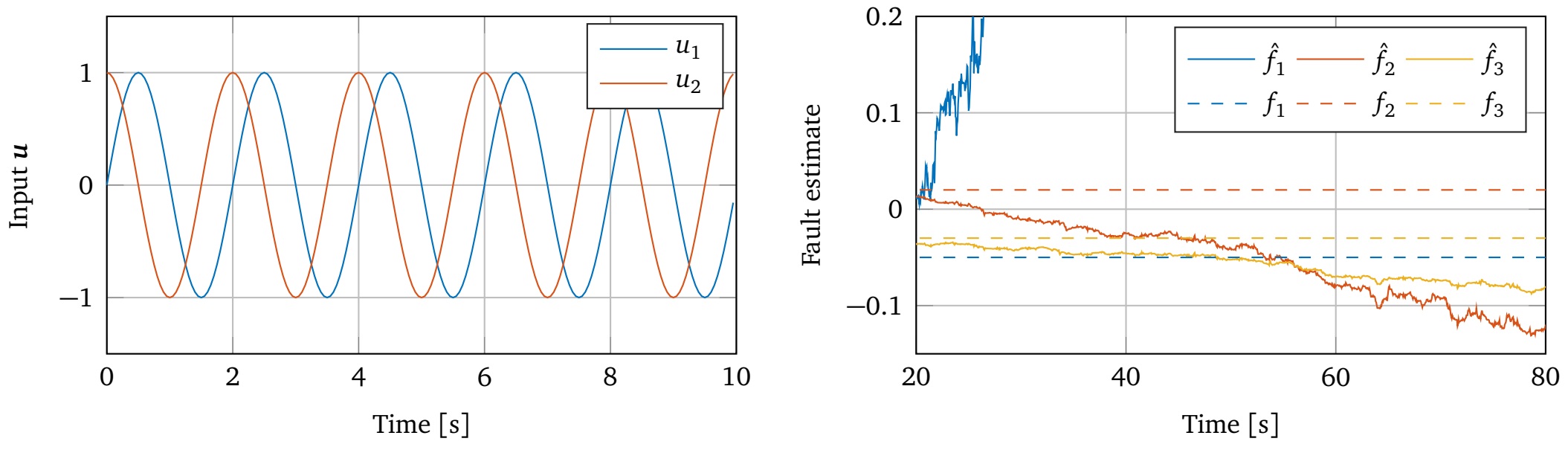

Figure 3: Input (left) and fault estimation results (right) for the second scenario, with simple sinusoidal input and noise.

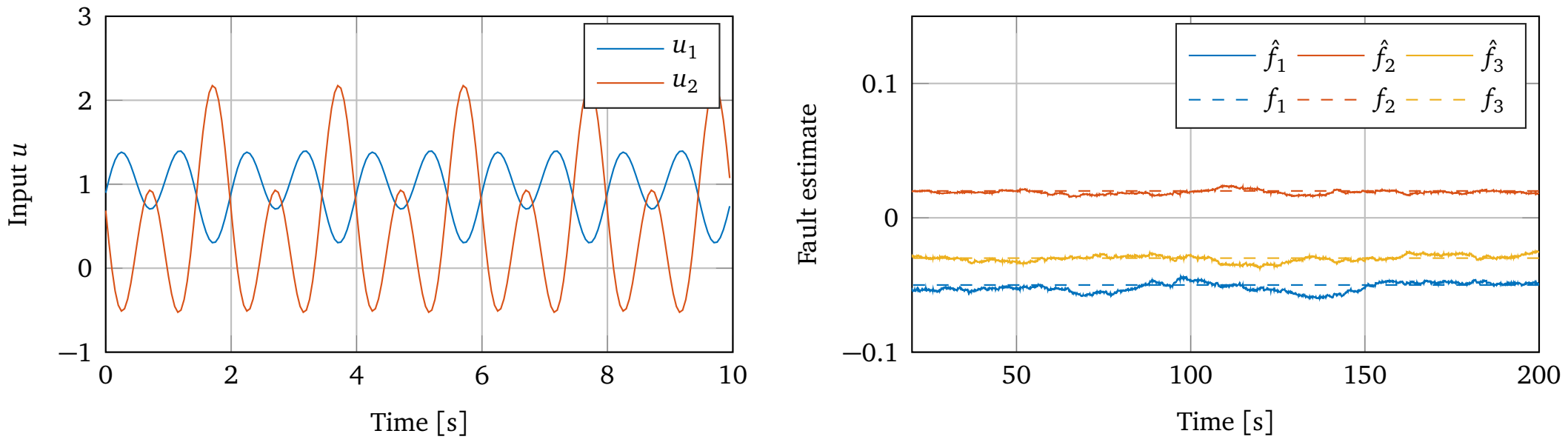

Figure 4: Input (left) and fault estimation results (right) for the third scenario, with optimized input and noise.

$\boldsymbol{H}_i'(\mathrm{q}) \not\equiv \mathbf{0}$, but it is still relatively small: $|\hat{\boldsymbol{f}}(t_f) - \boldsymbol{f}|/|\boldsymbol{f}| = 0.2$, where $t_f$ is the final time of the experiment. The performance increases substantially with the optimal input, where the final relative error drops to 0.0134. This is much smaller than the bound of 36.47 obtained by (13a), which indicates that in practice the estimation bias can be much lower than the conservative bound we have devised. In comparison to the noiseless case, it is striking to observe that the accuracy of the estimation is completely unacceptable once noise is introduced (Fig. 3). This is mainly due to the bad excitation properties of the input signal, as illustrated by the estimation performance using the optimized input in Fig. 4, as well as Fig. 5, which shows the minimal singular values of the regressors for both the sinusoidal and optimized inputs. The mean Euclidean norm of the estimation error, using the optimized input, is only about 3.4% of the norm of the fault vector. The mean squared error over the last half of the simulation is $3.3 \cdot 10^{-5}$, orders of magnitude smaller than the bound on the expected squared error given by the first-order-approximation result in Theorem 3.2, which is $5.0 \cdot 10^{-2}$. Finally, to assess the validity of the first-order-approximation, we computed the upper bound for the inverse SNR in (14), obtaining 0.12. Since it is significantly smaller than 1, the first order approximation is deemed accurate.

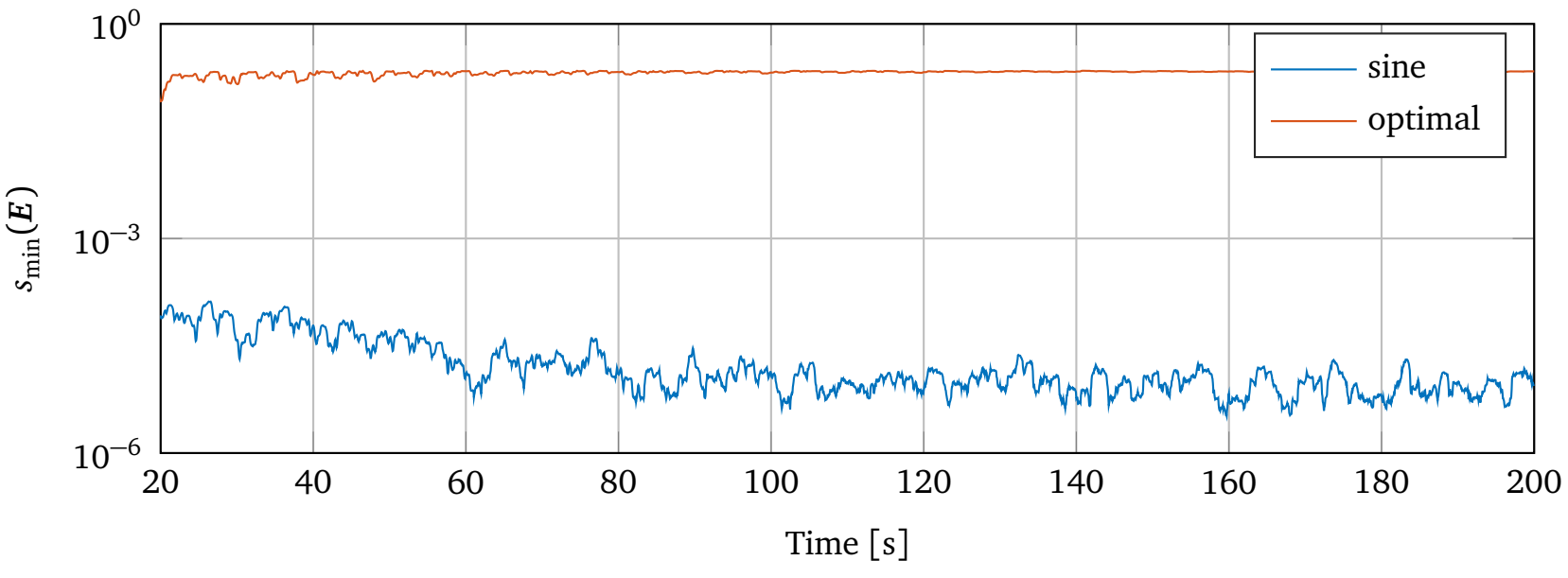

Figure 5: Running minimal singular values of $\boldsymbol{E}^h_{[t-20,t]}$ for the passive sinusoidal input and the optimized input.

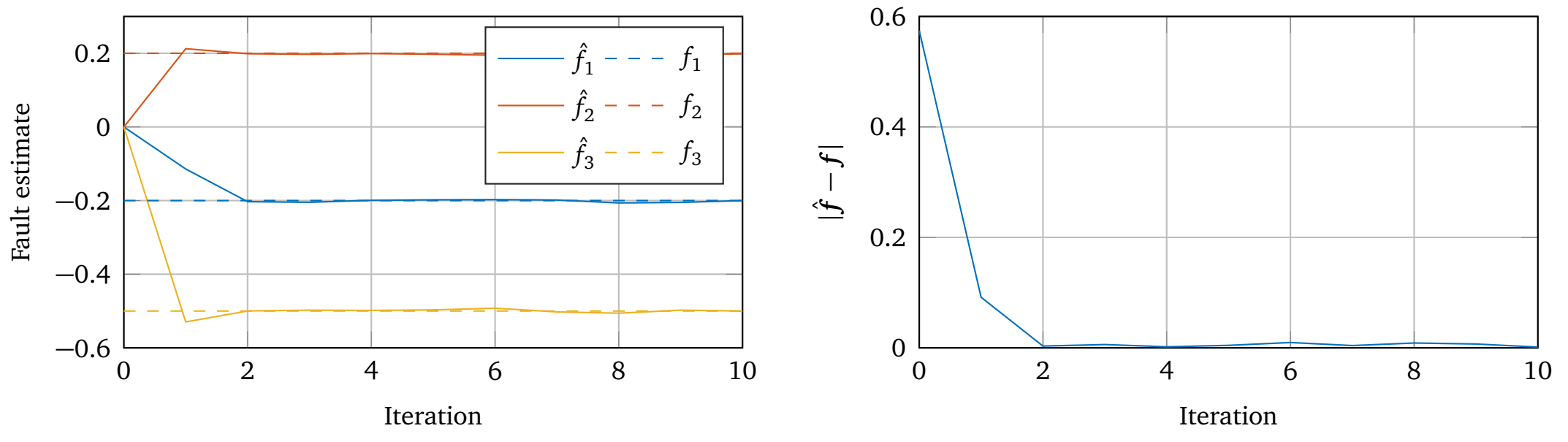

Figure 6: Fault estimation results (left) and estimation error norm (right) for the large fault scenario of Section 5.3.2. Each iteration corresponds to the last estimate after 40 seconds from the last update in the fault estimation filters.

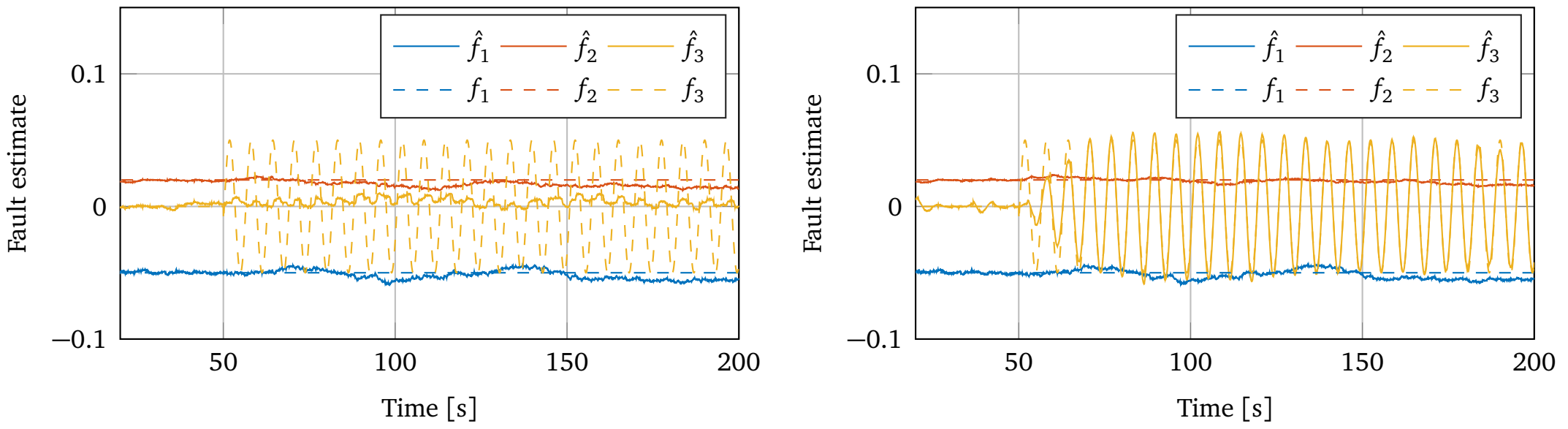

Figure 7: Fault estimation in the time-varying fault case using an estimator that assumes constant faults (left) and the estimation designed for the appropriate time-varying fault signature (right).

### 5.3.2 Large faults

In the previous subsection, we have seen the performance of our proposed fault estimation scheme for small faults. Now we present the results for significantly larger faults. In this case, the faults are constant with $f_1 = -0.2, f_2 = 0.2, f_3 = -0.5$ and we have followed the procedure suggested in Remark 3.2. The updates on the residual and regressor generators were made every 40 seconds. The same optimal input from Section 5.3.1 was used throughout the simulation. The resulting fault estimation performance is shown in Fig. 6; one may note that convergence is attained after two iterations only, after which the noise effects dominate the estimate variation.

### 5.3.3 Time-varying fault

In this last set, we simulate one time-varying fault: $f_1(t) \equiv -0.05, f_2(t) \equiv 0.02, f_3(t) = 0.05\sin(t)$. Following Section 3.3, we compute $\boldsymbol{G}'_{3,j}(\mathfrak{q})$. Since in our case study $\boldsymbol{G}_3(\mathfrak{q})$ is zero-order, we have only one regressor generator associated with $f_3$ and $\boldsymbol{G}'_{3,1} = \boldsymbol{G}_3$. Its corresponding input signal is $\boldsymbol{z}\sin(t)$. The optimal input designed in Section 5.3.1 is used in this experiment, and noise variance is kept at 1. Figure 7 shows the difference in fault estimation performance between using the exact same estimator as in Section 5.3.1, i.e., assuming all faults are constant, and using the correct time-varying estimator. Clearly, the performance is far superior with the correct assumption of time-varying fault for estimating $f_3$; in addition, a small improvement in the estimation of $f_2$ can be observed.

## 6 Conclusion

We have presented a fault estimation scheme for multiple multiplicative faults using a DAE framework, as well as numerical methods to compute its parameters. The focus is on small parametric faults, which have small effect in the residual signal, but allow for efficient estimation using a least-squares approach. The fault estimation performance is shown to be highly dependent on the richness of the involved signals, in particular the minimal singular value of a windowed trajectory of the regressor signal, which is highly dependent on the input signal of the plant. This motivated an active estimation approach, by means of an optimal periodic input design, for which we have provided a formulation and an algorithm for its efficient computation. The numerical example, despite being simple, highlights the importance of a properly designed input signal to enable accurate estimation in noisy conditions.

A prototype of the fault estimation method presented here has been successfully implemented in a larger scale realistic use case based on a chip manufacturing machine [35], with a 20-dimensional state space and 8 simultaneous faults. As part of the next steps, we are currently working on the implementation on a more realistic, nonlinear model of this system.

As a potential side contribution of this work, we believe our optimization algorithm can have an impact in other input design applications, such as system identification, active learning, and parameter estimation. There is evidence that the optimization problem (19), despite being non-convex, is well-behaved in the sense that either it does not contain spurious local maxima, or that this is the case with high probability as the problem dimensions increase.

# 7 Technical proofs

## 7.1 Proof of Theorem 3.1

*Proof of Lemma 3.1:* We start by pre-multiplying the system equation (1) with $\boldsymbol{N}(\mathsf{q})(\mathbf{I}-\sum_i f_i \boldsymbol{H}_i'(\mathsf{q})\boldsymbol{H}^\dagger(\mathsf{q}))$, which is a first-order approximation of the left null space of the system. To simplify notation, we drop the dependency of matrices on the operator $\mathsf{q}$.

$$\begin{aligned}\boldsymbol{N}\left(\mathbf{I}-\sum_i f_i\boldsymbol{H}_i'\boldsymbol{H}^\dagger\right)\left(\boldsymbol{H}+\sum_i f_i\boldsymbol{H}_i'\right)\xi \\ +\boldsymbol{N}\left(\mathbf{I}-\sum_i f_i\boldsymbol{H}_i'\boldsymbol{H}^\dagger\right)\left(\boldsymbol{L}+\sum_i f_i\boldsymbol{L}_i'\right)\boldsymbol{z} \\ +\boldsymbol{N}\left(\mathbf{I}-\sum_i f_i\boldsymbol{H}_i'\boldsymbol{H}^\dagger\right)\boldsymbol{W}\boldsymbol{w}=\mathbf{0}.\end{aligned}$$

Expanding, we get

$$\begin{aligned}\left(\boldsymbol{N}\boldsymbol{H}+\sum_i f_i\boldsymbol{N}\boldsymbol{H}_i'-\sum_i f_i\boldsymbol{N}\boldsymbol{H}_i'-\sum_i\sum_j f_if_j\boldsymbol{N}\boldsymbol{H}_i'\boldsymbol{H}^\dagger\boldsymbol{H}_j'\right)\xi \\ +\left(\boldsymbol{N}\boldsymbol{L}+\sum_i f_i\boldsymbol{N}\boldsymbol{L}_i'-\sum_i f_i\boldsymbol{N}\boldsymbol{H}_i'\boldsymbol{H}^\dagger\boldsymbol{L}-\sum_i\sum_j f_if_j\boldsymbol{N}\boldsymbol{H}_i'\boldsymbol{H}^\dagger\boldsymbol{L}_j'\right)\boldsymbol{z} \\ +\boldsymbol{N}\left(\mathbf{I}-\sum_i f_i\boldsymbol{H}_i'\boldsymbol{H}^\dagger\right)\boldsymbol{W}\boldsymbol{w}=0.\end{aligned}$$

Since $\boldsymbol{N}\boldsymbol{H}=\mathbf{0}$, and $\boldsymbol{N}\boldsymbol{L}\boldsymbol{z}=\boldsymbol{d}(\mathsf{q})r$, we obtain

$$\begin{aligned}-\left(\sum_i\sum_j f_if_j\boldsymbol{N}\boldsymbol{H}_i'\boldsymbol{H}^\dagger\boldsymbol{H}_j'\right)\xi+\boldsymbol{d}(\mathsf{q})r+\left(\sum_i f_i\boldsymbol{N}\boldsymbol{L}_i'-\sum_i f_i\boldsymbol{N}\boldsymbol{H}_i'\boldsymbol{H}^\dagger\boldsymbol{L}\right)\boldsymbol{z} \\ -\left(\sum_i\sum_j f_if_j\boldsymbol{N}\boldsymbol{H}_i'\boldsymbol{H}^\dagger\boldsymbol{L}_j'\right)z-\boldsymbol{N}\left(\sum_i f_i\boldsymbol{H}_i'\boldsymbol{H}^\dagger\right)\boldsymbol{W}\boldsymbol{w}+\boldsymbol{N}\boldsymbol{W}\boldsymbol{w}=0.\end{aligned}$$

Finally, let $\boldsymbol{J}_{i,j} := \begin{bmatrix} -\boldsymbol{N}\boldsymbol{H}_i'\boldsymbol{H}^\dagger\boldsymbol{H}_j' & -\boldsymbol{N}\boldsymbol{H}_i'\boldsymbol{H}^\dagger\boldsymbol{L}_j' \end{bmatrix}$ and $\boldsymbol{F}_i := -\boldsymbol{N}\boldsymbol{H}_i'\boldsymbol{H}^\dagger\boldsymbol{W}$, which are identically zero if $\boldsymbol{H}_i' \equiv \boldsymbol{0}$ for all $i$. We arrive at the desired simplified expression:

$$\boldsymbol{d}(\mathrm{q})r = \left(\sum_i f_i\boldsymbol{N}\boldsymbol{H}_i'\boldsymbol{H}^\dagger\boldsymbol{L} - f_i\boldsymbol{N}\boldsymbol{L}_i'\right)\boldsymbol{z} + \sum_{ij} f_i f_j \boldsymbol{J}_{i,j}(\mathrm{q})\begin{bmatrix}\xi\\ \boldsymbol{z}\end{bmatrix} + \sum_i \mathcal{O}(f_i)\boldsymbol{F}_i(\mathrm{q})\boldsymbol{w} - \boldsymbol{N}\boldsymbol{W}\boldsymbol{w}.$$

■

Now we provide a useful lemma regarding a notion of *symbolic rank* of polynomial matrices.

**Lemma 7.1.** *Consider a polynomial matrix $\boldsymbol{H} \in \mathbb{R}[\mathrm{q}]^{m\times n}$ of degree $d$. The following statements hold:*

*(i) There exists a nonzero $\boldsymbol{v} \in \mathbb{R}^n$ such that $\boldsymbol{H}(\mathrm{q})\boldsymbol{v} \equiv \boldsymbol{0}$ if and only if* blkrow$(\boldsymbol{H}(\mathrm{q})^\mathsf{T})$ *is row-rank deficient;*

*(ii) There exists a nonzero $\boldsymbol{v} \in \mathbb{R}^m$ such that $\boldsymbol{v}^\mathsf{T}\boldsymbol{H}(\mathrm{q}) \equiv \boldsymbol{0}$ if and only if* blkrow$(\boldsymbol{H}(\mathrm{q}))$ *is row-rank deficient.*

*Proof:* We prove item (ii), which immediately gives (i) by transposing the matrices. First, we start by noticing that $\exists \boldsymbol{v} \in \mathbb{R}^n : \boldsymbol{v}^\mathsf{T}\boldsymbol{H}(\mathrm{q}) \equiv \boldsymbol{0}$ iff $\exists \boldsymbol{v} \in \mathbb{R}^n : \boldsymbol{v}^\mathsf{T}\boldsymbol{H}_i = \boldsymbol{0}$ for all $i = 0, 1, \ldots, d$. This, in turn, is equivalent to $\exists \boldsymbol{v} \in \mathbb{R}^n : \boldsymbol{v}^\mathsf{T}\text{blkrow}(\boldsymbol{H}(\mathrm{q})) = \boldsymbol{0}$, which is the definition of row-rank deficiency. ■

The following Lemma shows that full symbolic rank of a filter gives linearly independent signals provided its input is persistently exciting:

**Lemma 7.2.** *Consider a polynomial matrix $\boldsymbol{H} \in \mathbb{R}[\mathrm{q}]^{m\times n}$ of degree $d$, where $\mathrm{q}$ represents a differential or time-shift operator. Assume* blkrow$(\boldsymbol{H}(\mathrm{q}))$ *is full-row-rank. Then, if $\boldsymbol{z}$ is PE of order $d+1$, the output of $\boldsymbol{H}(\mathrm{q})\boldsymbol{z}$ is linearly independent.*

*Proof:* Suppose $\boldsymbol{H}(\mathrm{q})\boldsymbol{z}$ has a linear dependence. Then there exists a nonzero $\boldsymbol{v} \in \mathbb{R}^m$ such that $\boldsymbol{v}^\mathsf{T}\boldsymbol{H}(\mathrm{q})\boldsymbol{z}(t) = 0$ for all $t$. By Lemma 7.1, for all nonzero $\boldsymbol{v} \in \mathbb{R}^m$, it holds that $\boldsymbol{v}^\mathsf{T}\boldsymbol{H}(\mathrm{q}) \not\equiv \boldsymbol{0}$. Thus, $\boldsymbol{a}(\mathrm{q})^\mathsf{T} := \boldsymbol{v}^\mathsf{T}\boldsymbol{H}(\mathrm{q})$ is a polynomial vector of degree less than or equal to $d$. Hence, $\boldsymbol{a}(\mathrm{q})^\mathsf{T}\boldsymbol{z}(t) = 0$ for all $t$, which can be rewritten as the RHS of (4), contradicting the fact that $\boldsymbol{z}$ is PE of order $d+1$. ■

Lemmas 3.1, 7.1 and 7.2 allow us to prove Theorem 3.1.

*Proof of Theorem 3.1:* By Lemma 2.1, $\boldsymbol{N}(\mathrm{q})\boldsymbol{H}(\mathrm{q}) \equiv \boldsymbol{0}$ is equivalent to (7a), and equations (6) and (8) are equivalent. From Lemma 3.1, if Assumption 2.1 holds, then taking $\boldsymbol{N}(\mathrm{q})$ s.t. $\boldsymbol{N}(\mathrm{q})\boldsymbol{H}(\mathrm{q}) \equiv \boldsymbol{0}$ and $\boldsymbol{M}(\mathrm{q})$ according to (6) yields (5). For small $\boldsymbol{f}$, and taking $\boldsymbol{T}(\mathrm{q}) := -\boldsymbol{N}(\mathrm{q})\boldsymbol{W}(\mathrm{q})$, we obtain the approximation (3), fulfilling requirement (b) of Definition 3.2.

Finally, Lemmas 7.1 and 7.2 establish that (7b) implies that requirements (a) and (c) of Definition 3.2 are satisfied. This concludes the proof. ■

## 7.2 Proof of Theorem 3.2

Let us recall the expression for the first-order approximation of the fault estimation error (12):

$$\breve{\boldsymbol{f}} - \boldsymbol{f} = \boldsymbol{E}^\dagger\boldsymbol{R}_{\mathrm{NL}} - \boldsymbol{E}^\dagger(\boldsymbol{R}_w - \boldsymbol{E}_w\boldsymbol{f}) - \left((\boldsymbol{E}^\mathsf{T}\boldsymbol{E})^{-1}\boldsymbol{E}_w^\mathsf{T}\boldsymbol{P}_\perp - \boldsymbol{E}^\dagger\boldsymbol{E}_w\boldsymbol{E}^\dagger\right)\boldsymbol{R}_{\mathrm{NL}}.$$

We divide the proof of Theorem 3.2 into two parts: the proof of its bias bound expression (13a), and its total variance bound expression (13b).

**Proof of** (13a)**:** By assumption, $\boldsymbol{R}_w$ and $\boldsymbol{E}_w$ are zero-mean, and as such, we can state that $\mathbb{E}(\breve{\boldsymbol{f}} - \boldsymbol{f}) = \boldsymbol{E}^\dagger\boldsymbol{R}_{\mathrm{NL}}$. Thus, its expression can be obtained with $\boldsymbol{w} \equiv \boldsymbol{0}$. In this case, for any $t$, (5) gives

$$r(t) = \boldsymbol{f}^\mathsf{T}\boldsymbol{e}(t) + \sum_{i,j} f_i f_j \mathcal{T}_{i,j}^J(\mathfrak{z})\begin{bmatrix}\xi(t)\\ \boldsymbol{z}(t)\end{bmatrix}. \tag{21}$$

Hence,

$$\left|r(t) - \boldsymbol{f}^\mathsf{T}\boldsymbol{e}(t)\right| \leq \sum_{i,j} |f_i f_j| \left\|\mathcal{T}_{i,j}^J\right\|_{\infty,\infty} \left\|\begin{bmatrix}\xi\\ \boldsymbol{z}\end{bmatrix}\right\|_\infty$$

where $\|\mathscr{T}\|_{\infty,\infty}$ is the peak-to-peak gain of $\mathscr{T}$. Denote by $C := \max_{i,j} \left\|\mathscr{T}_{i,j}^J\right\|_{\infty,\infty} \left\|\begin{bmatrix}\xi\\ \boldsymbol{z}\end{bmatrix}\right\|_{\infty}$. We then have

$$\left|r(t) - \boldsymbol{f}^\mathsf{T}\boldsymbol{e}(t)\right| \leq C\sum_{i,j}|f_i f_j| \leq Cm\boldsymbol{f}^\mathsf{T}\boldsymbol{f}, \tag{22}$$

where the last inequality follows from Young's inequality. Let $b := Cm\boldsymbol{f}^\mathsf{T}\boldsymbol{f}$. Then,

$$-b \leq r(t) - \boldsymbol{f}^\mathsf{T}\boldsymbol{e}(t) \leq b \implies -b\mathbf{1} \leq \boldsymbol{R} - \boldsymbol{E}\boldsymbol{f} \leq b\mathbf{1},$$

where $\mathbf{1}$ is a vector of ones and the inequality holds element-wise. Hence,

$$|\boldsymbol{R} - \boldsymbol{E}\boldsymbol{f}| \leq b\sqrt{N}. \tag{23}$$

Now, $\hat{\boldsymbol{f}}(t) - \boldsymbol{f} = \boldsymbol{E}^\dagger(\boldsymbol{R} - \boldsymbol{E}\boldsymbol{f})$. Thus,

$$\left|\hat{\boldsymbol{f}}(t) - \boldsymbol{f}\right| = \left|\boldsymbol{E}^\dagger(\boldsymbol{R} - \boldsymbol{E}\boldsymbol{f})\right| \leq \left\|\boldsymbol{E}^\dagger\right\|_2 |\boldsymbol{R} - \boldsymbol{E}\boldsymbol{f}| \leq s_{\min}(\boldsymbol{E})^{-1} b\sqrt{N}, \tag{24}$$

where $\left\|\boldsymbol{E}^\dagger\right\|_2 = s_{\min}(\boldsymbol{E})^{-1}$ comes from the fact that the singular values of $\boldsymbol{A}^\dagger$ are the reciprocals of the singular values of $\boldsymbol{A}$ for any $\boldsymbol{A}$.

Since $\boldsymbol{E}$ is full rank, $s_{\min}(\boldsymbol{E})^{-1} < \infty$, and thus the proof is concluded by taking $A = \max_{i,j} \left\|\mathscr{T}_{i,j}^J\right\|_{\infty,\infty}$. ■

To prove Theorem 3.2, we shall use the notion of stochastic norm presented by [30]. For a random matrix $\boldsymbol{A}$, its stochastic norm is defined by $\|\boldsymbol{A}\|_\mathrm{S}^2 = \mathbb{E}(\|\boldsymbol{A}\|_\mathrm{F}^2)$. We begin with some useful bounds.

**Lemma 7.3.** *Let $\boldsymbol{y} = \mathscr{T}(\mathfrak{z})\boldsymbol{w}$ be a discrete-time $m \times n$ LTI w $\mathscr{H}_\infty^\mathrm{F}$ norm of $\mathscr{T}$ is $\eta < \infty$ and where $w_i$ are independent zero-mean i.i.d. random processes with variance $\sigma^2$. Let $\boldsymbol{\Sigma} \in \mathbb{S}_+^N$ be the auto-covariance matrix defined by $(\boldsymbol{\Sigma})_{i,j} = \mathbb{E}[\boldsymbol{y}(t)^\mathsf{T}\boldsymbol{y}(t - |i-j+1|)]$. Then, uniformly in $N$, it holds that $\lambda_{\max}(\boldsymbol{\Sigma}) \leq \eta^2\sigma^2$.*

*Proof:* First, we decompose the auto-covariance of $\boldsymbol{y}$

$$\begin{aligned}
\mathbb{E}[\boldsymbol{y}(t)^\mathsf{T}\boldsymbol{y}(t-\tau)] &= \mathbb{E}\left[\sum_{i=1}^m \left(\sum_{j=1}^n w_j(t) * h_{i,j}(t)\right)^\mathsf{T} \left(\sum_{k=1}^n w_k(t-\tau) * \boldsymbol{h}_{i,k}(t-\tau)\right)\right] \\
&= \mathbb{E}\left[\sum_{i=1}^n\sum_{j=1}^m (w_j(t) * h_{i,j}(t))(w_j(t-\tau) * h_{i,j}(t-\tau))\right] \\
&= \sum_{i=1}^n\sum_{j=1}^m \mathbb{E}\left[(w_j(t) * h_{i,j}(t))(w_j(t-\tau) * h_{i,j}(t-\tau))\right],
\end{aligned}$$

where $h_{i,j}$ is the impulse response of the entry $\mathscr{T}_{i,j}$, and we have used independence, i.e., $i \neq j \implies \mathbb{E}[w_i(t)w_j(t')] = 0$ for all $t, t'$, in the third equation. We conclude that $\boldsymbol{\Sigma} = \sum_{i=1}^n\sum_{j=1}^m \boldsymbol{\Sigma}_{i,j}$, where $\boldsymbol{\Sigma}_{i,j}$ is the auto-covariance matrix of $\mathscr{T}_{i,j}w_i$.

Now, consider the auto-covariance function $\rho_{i,j}$ of $\mathscr{T}_{i,j}(\mathfrak{z})w_i$. According to [36, Chap. 10], the discrete-time Fourier transform $\varsigma_{i,j}$ of $\rho_{i,j}$ satisfies $\varsigma_{i,j}(\mathrm{j}\omega) = \sigma^2|\mathscr{T}_{i,j}(\exp(\mathrm{j}\omega))|^2$, where we use the fact that the Fourier transform of the auto-covariance of $w_i$ is constant and equal to $\sigma^2$. By [37, Lemma 4.1], a Toeplitz matrix $\boldsymbol{\Sigma} \in \mathbb{S}_+^T$ satisfies, uniformly in $N$, $\lambda_{\max}(\boldsymbol{\Sigma}_T) \leq \operatorname{ess\,sup}_{\omega\in[0,2\pi]} \varsigma(\exp(\mathrm{j}\omega))$. Hence,

$$\lambda_{\max}(\boldsymbol{\Sigma}_T) \leq \sigma^2 \operatorname*{ess\,sup}_{\omega\in[0,2\pi]} \sum_{i=1}^{n_\mathrm{w}} |\mathscr{T}_i(\exp(\mathrm{j}\omega))|^2 = \sigma^2\eta^2,$$

by definition of $\mathscr{H}_\infty^\mathrm{F}$ norm. ■

**Lemma 7.4.** *Let $\gamma_F$ ($\eta_F$) and $\gamma_W$ ($\eta_W$) be the $\mathscr{H}_2$ ($\mathscr{H}_\infty^{\mathrm{F}}$) norms of the transfer functions $\mathscr{T}^F$ and $\mathscr{T}^W$. Then,*

$$\mathbb{E}[\mathrm{tr}(\boldsymbol{E}_w^\mathsf{T}\boldsymbol{E}_w)] \leq \sigma^2 N \gamma_F^2 \tag{25}$$

$$\lambda_{\max}(\mathrm{Var}(\boldsymbol{R}_w - \boldsymbol{E}_w\boldsymbol{f})) \leq (|\boldsymbol{f}|^2+1)\sigma^2\eta^2, \tag{26}$$

$$\mathrm{tr}\,\mathrm{Var}(\boldsymbol{E}_w^\mathsf{T}\boldsymbol{b}) \leq |\boldsymbol{b}|^2\sigma^2\eta_F^2, \tag{27}$$

*where $\gamma^2 := \gamma_F^2 + \gamma_W^2$, and $\eta^2 = \eta_F^2 + \eta_W^2$*

*Proof:* Recall that the $\mathscr{H}_2$ norm $\gamma$ of an LTI system $\boldsymbol{y} = \mathscr{T}(\mathfrak{z})\boldsymbol{w}$ satisfies

$$\lim_{N\to\infty} \frac{1}{N}\sum_{t=0}^{N} \mathbb{E}[\boldsymbol{y}(t)^\mathsf{T}\boldsymbol{y}(t)] = \lim_{N\to\infty} \frac{1}{N}\sum_{t=0}^{N} \mathrm{tr}\,\mathrm{Var}(\boldsymbol{y}(t)) = \gamma^2.$$

when $\boldsymbol{w}$ is an i.i.d. random process with variance $\mathbf{I}$.

Because the variance $\mathbb{E}[\boldsymbol{y}(t)^\mathsf{T}\boldsymbol{y}(t)]$ is monotonically increasing with $t$, this implies that for any $t_1$,

$$\sum_{t=t_1}^{t_1+N} \mathbb{E}[\boldsymbol{y}(t)^\mathsf{T}\boldsymbol{y}(t)] \leq N\gamma^2. \tag{28}$$

Now, $\mathbb{E}[\mathrm{tr}(\boldsymbol{E}_w^\mathsf{T}\boldsymbol{E}_w)] = \mathbb{E}\left[\sum_{i=1}^m (\boldsymbol{E}_w^i)^\mathsf{T}\boldsymbol{E}_w^i\right] = \sum_{i=1}^m \mathbb{E}[(\boldsymbol{E}_w^i)^\mathsf{T}\boldsymbol{E}_w^i] = \sum_{t-(N-1)}^{t} \boldsymbol{e}(t)^\mathsf{T}\boldsymbol{e}(t)$, where $\boldsymbol{E}_w^i$ is the $i$-th column of $\boldsymbol{E}_w$. Applying (28), while recalling that $\mathrm{Var}(\boldsymbol{w}) = \sigma^2\mathbf{I}$, leads to (25).

For $\lambda_{\max}(\mathrm{Var}(\boldsymbol{R}_w - \boldsymbol{E}_w\boldsymbol{f}))$, notice that $r(t) - \boldsymbol{e}(t)^\mathsf{T}\boldsymbol{f}$ is the output of the system $\mathscr{T}' := \mathscr{T}^W - \sum_{i=1}^m f_i\mathscr{T}_i^F$. By the triangle inequality, its $\mathscr{H}_\infty^{\mathrm{F}}$ norm $\eta'$ satisfies $\eta' \leq \eta^W + |\boldsymbol{f}|\,\eta^F \leq \sqrt{(1+|\boldsymbol{f}|^2)\eta^2}$ (by Cauchy–Schwarz). The bound (26) is then obtained by applying Lemma 7.3.

For the last bound,

$$\mathrm{tr}\left(\mathrm{Var}(\boldsymbol{E}_w^\mathsf{T}\boldsymbol{b})\right) = \sum_{i=1}^m \mathrm{Var}(\boldsymbol{b}^\mathsf{T}\boldsymbol{E}_w^i) = \sum_{i=1}^m \boldsymbol{b}^\mathsf{T}\boldsymbol{\Sigma}_i\boldsymbol{b} = \boldsymbol{b}^\mathsf{T}\left(\sum_{i=1}^m \boldsymbol{\Sigma}_i\right)\boldsymbol{b} = \boldsymbol{b}^\mathsf{T}\boldsymbol{\Sigma}\boldsymbol{b}, \tag{29}$$

where $\boldsymbol{\Sigma}_i$ is the Toeplitz matrix composed by the auto-covariances of $e_i(t)$, and $\boldsymbol{\Sigma}$ is the auto-covariance matrix defined by $\mathbb{E}[\boldsymbol{e}(t)^\mathsf{T}\boldsymbol{e}(t-\tau)]$ as in Lemma 7.3. Hence, this Lemma leads to

$$\mathrm{tr}\,\mathrm{Var}(\boldsymbol{E}_w^\mathsf{T}\boldsymbol{b}) \leq |\boldsymbol{b}|^2\,\lambda_{\max}(\boldsymbol{\Sigma}) \leq |\boldsymbol{b}|^2\,\sigma^2\eta_F^2.$$

■

The following fact is taken from [38, Prop. 8.4.13]:

**Lemma 7.5.** *For any two p.s.d. matrices $\boldsymbol{A}, \boldsymbol{B} \in \mathbb{S}_+^n$, it holds that* $\mathrm{tr}(\boldsymbol{A}\boldsymbol{B}) \leq \mathrm{tr}(\boldsymbol{A})\lambda_{\max}(\boldsymbol{B})$.

**Proof of** (13b): First, notice that the stochastic norm of the estimation error satisfies $\left\|\breve{\boldsymbol{f}} - \boldsymbol{f}\right\|_{\mathrm{S}}^2 = \mathbb{E}(\breve{\boldsymbol{f}} - \boldsymbol{f})^2 + \mathrm{tr}(\mathrm{Var}(\breve{\boldsymbol{f}} - \boldsymbol{f}))$. Hence, we obtain

$$\mathrm{tr}(\mathrm{Var}(\breve{\boldsymbol{f}} - \boldsymbol{f})) = \left\|\boldsymbol{E}^\dagger(\boldsymbol{R}_w - \boldsymbol{E}_w\boldsymbol{f}) + \left((\boldsymbol{E}^\mathsf{T}\boldsymbol{E})^{-1}\boldsymbol{E}_w^\mathsf{T}\boldsymbol{P}_\perp - \boldsymbol{E}^\dagger\boldsymbol{E}_w\boldsymbol{E}^\dagger\right)\boldsymbol{R}_{\mathrm{NL}}\right\|_{\mathrm{S}}^2.$$

Since $\boldsymbol{E}^\dagger\boldsymbol{P}_\perp = \mathbf{0}$, we have, by [30, Theorem 2.4],

$$\begin{aligned}\mathrm{tr}(\mathrm{Var}(\breve{\boldsymbol{f}} - \boldsymbol{f})) &= \left\|\boldsymbol{E}^\dagger(\boldsymbol{R}_w - \boldsymbol{E}_w\boldsymbol{f}) - \boldsymbol{E}^\dagger\boldsymbol{E}_w\boldsymbol{E}^\dagger\boldsymbol{R}_{\mathrm{NL}}\right\|_{\mathrm{S}}^2 + \left\|(\boldsymbol{E}^\mathsf{T}\boldsymbol{E})^{-1}\boldsymbol{E}_w^\mathsf{T}\boldsymbol{P}_\perp\boldsymbol{R}_{\mathrm{NL}}\right\|_{\mathrm{S}}^2 \\ &\leq 2\left\|\boldsymbol{E}^\dagger(\boldsymbol{R}_w - \boldsymbol{E}_w\boldsymbol{f})\right\|_{\mathrm{S}}^2 + 2\left\|\boldsymbol{E}^\dagger\boldsymbol{E}_w\boldsymbol{E}^\dagger\boldsymbol{R}_{\mathrm{NL}}\right\|_{\mathrm{S}}^2 + \left\|(\boldsymbol{E}^\mathsf{T}\boldsymbol{E})^{-1}\boldsymbol{E}_w^\mathsf{T}\boldsymbol{P}_\perp\boldsymbol{R}_{\mathrm{NL}}\right\|_{\mathrm{S}}^2,\end{aligned} \tag{30}$$

where we have used linearity of expectation and the fact that $\left(\sum_{i=1}^N f_i\right)^2 \leq N\sum_{i=1}^N f_i^2$. Now we bound each term, recalling the following facts:

F1. For any matrix $\boldsymbol{A}$, the eigenvalues of $\boldsymbol{A}^{\mathsf{T}}\boldsymbol{A}$ are the squared singular values of $\boldsymbol{A}$, thus it holds for any $p \in \mathbb{R}$ that $\operatorname{tr}((\boldsymbol{A}^{\mathsf{T}}\boldsymbol{A})^p) = \sum_i s_i^{2p}(\boldsymbol{A})$.

F2. For a random variable $\boldsymbol{x}$ with mean $\boldsymbol{m}$ and covariance matrix $\boldsymbol{\Sigma}$, it holds that [39] $\mathbb{E}[\boldsymbol{x}^{\mathsf{T}}\boldsymbol{A}\boldsymbol{x}] = \operatorname{tr}(\boldsymbol{A}\boldsymbol{\Sigma}) + \boldsymbol{m}^{\mathsf{T}}\boldsymbol{A}\boldsymbol{m}$. Hence, for a zero-mean $\boldsymbol{x}, \mathbb{E}[|\boldsymbol{A}\boldsymbol{x}|_2^2] = \operatorname{tr}(\boldsymbol{A}^{\mathsf{T}}\boldsymbol{A}\Sigma)$ and $\operatorname{Var}(\boldsymbol{f}^{\mathsf{T}}\boldsymbol{x}) = \boldsymbol{f}^{\mathsf{T}}\boldsymbol{\Sigma}\boldsymbol{f}$.

Bounding the first term in (30), we obtain

$$\begin{aligned}
\left\|\boldsymbol{E}^{\dagger}(\boldsymbol{R}_w - \boldsymbol{E}_w\boldsymbol{f})\right\|_{\mathrm{S}}^2 &= \mathbb{E}\left[\left|\boldsymbol{E}^{\dagger}(\boldsymbol{R}_w - \boldsymbol{E}_w\boldsymbol{f})\right|^2\right] \underset{\mathbb{E}[\boldsymbol{R}_w - \boldsymbol{E}_w\boldsymbol{f}]=\boldsymbol{0}}{\overset{\text{F2}}{=}} \operatorname{tr}(\boldsymbol{E}^{\dagger\mathsf{T}}\boldsymbol{E}^{\dagger}\operatorname{Var}(\boldsymbol{R}_w - \boldsymbol{E}_w\boldsymbol{f})) \\
&\overset{\text{Lemma 7.5}}{\leq} \operatorname{tr}(\boldsymbol{E}^{\dagger\mathsf{T}}\boldsymbol{E}^{\dagger})\lambda_{\max}(\operatorname{Var}(\boldsymbol{R}_w - \boldsymbol{E}_w\boldsymbol{f})) = \operatorname{tr}(\boldsymbol{E}(\boldsymbol{E}^{\mathsf{T}}\boldsymbol{E})^{-1}(\boldsymbol{E}^{\mathsf{T}}\boldsymbol{E})^{-1}\boldsymbol{E}^{\mathsf{T}})\lambda_{\max}(\operatorname{Var}(\boldsymbol{R}_w - \boldsymbol{E}_w\boldsymbol{f})) \\
&= \operatorname{tr}(\boldsymbol{E}^{\mathsf{T}}\boldsymbol{E}(\boldsymbol{E}^{\mathsf{T}}\boldsymbol{E})^{-1}(\boldsymbol{E}^{\mathsf{T}}\boldsymbol{E})^{-1})\lambda_{\max}(\operatorname{Var}(\boldsymbol{R}_w - \boldsymbol{E}_w\boldsymbol{f})) = \operatorname{tr}((\boldsymbol{E}^{\mathsf{T}}\boldsymbol{E})^{-1})\lambda_{\max}(\operatorname{Var}(\boldsymbol{R}_w - \boldsymbol{E}_w\boldsymbol{f})) \\
&\overset{\text{F1}}{=} \sum_{i=1}^{m} s_i^{-2}\lambda_{\max}(\operatorname{Var}(\boldsymbol{R}_w - \boldsymbol{E}_w\boldsymbol{f})) \overset{\text{Eq. (26)}}{\leq} (|\boldsymbol{f}|^2 + 1)\sigma^2\eta^2\sum_{i=1}^{m} s_i^{-2}. \qquad (31)
\end{aligned}$$

For the second and third terms, first note that $\tilde{\boldsymbol{f}} := \boldsymbol{E}^{\dagger}\boldsymbol{R}_{\mathrm{NL}}$ is precisely the error in estimation for the noiseless case, whose Euclidean-norm upper bound has been provided in (13a). Expanding the second term,

$$\left\|\boldsymbol{E}^{\dagger}\boldsymbol{E}_w\boldsymbol{E}^{\dagger}\boldsymbol{R}_{\mathrm{NL}}\right\|_{\mathrm{S}}^2 = \left\|\boldsymbol{E}^{\dagger}\boldsymbol{E}_w\tilde{\boldsymbol{f}}\right\|_{\mathrm{S}}^2 \overset{\text{F1}}{=} \mathbb{E}\left[\left|\boldsymbol{E}^{\dagger}\boldsymbol{E}_w\tilde{\boldsymbol{f}}\right|^2\right] \leq \lambda_{\max}(\operatorname{Var}(\boldsymbol{E}_w\tilde{\boldsymbol{f}}))\sum_{i=1}^{m} s_i^{-2},$$

where the last inequality comes from following the same steps as in (31). Following the steps of the proof of (26) in Lemma 7.4, we get

$$\left\|\boldsymbol{E}^{\dagger}\boldsymbol{E}_w\boldsymbol{E}^{\dagger}\boldsymbol{R}_{\mathrm{NL}}\right\|_{\mathrm{S}}^2 \leq \left|\tilde{\boldsymbol{f}}\right|^2\sigma^2\eta_F^2\sum_{i=1}^{m} s_i^{-2} \overset{\text{(13a)}}{=} B^2\sigma^2\eta_F^2\sum_{i=1}^{m} s_i^{-2}. \qquad (32)$$

For the third term, let $\boldsymbol{b} := \boldsymbol{P}_{\perp}\boldsymbol{R}_{\mathrm{NL}}$. We again follow similar steps to (31) to get

$$\begin{aligned}
\left\|(\boldsymbol{E}^{\mathsf{T}}\boldsymbol{E})^{-1}\boldsymbol{E}_w^{\mathsf{T}}\boldsymbol{b}\right\|_{\mathrm{S}}^2 &\leq \lambda_{\max}((\boldsymbol{E}^{\mathsf{T}}\boldsymbol{E})^{-2})\operatorname{tr}(\operatorname{Var}(\boldsymbol{E}_w^{\mathsf{T}}\boldsymbol{b})) = \operatorname{tr}(\operatorname{Var}(\boldsymbol{E}_w^{\mathsf{T}}\boldsymbol{b}))s_{\min}(\boldsymbol{E})^{-4} \\
&\overset{(27)}{\leq} |\boldsymbol{b}|^2\sigma^2\eta_F^2 s_{\min}(\boldsymbol{E})^{-4} \leq |\boldsymbol{b}|^2\sigma^2\eta_F^2 s_{\min}(\boldsymbol{E})^{-4}. \qquad (33)
\end{aligned}$$

Since $\boldsymbol{P}_{\perp}$ is a projection matrix, $|\boldsymbol{b}| \leq |\boldsymbol{R}_{\mathrm{NL}}|$. By the proof of (13a), $|\boldsymbol{R}_{\mathrm{NL}}| \leq Am|\boldsymbol{f}|^2\sqrt{N}\left\|\begin{bmatrix}\boldsymbol{\xi}^{\mathsf{T}} & \boldsymbol{z}^{\mathsf{T}}\end{bmatrix}\right\|_{\infty} = Bs_{\min}(\boldsymbol{E})$. Therefore,

$$\left\|(\boldsymbol{E}^{\mathsf{T}}\boldsymbol{E})^{-1}\boldsymbol{E}_w^{\mathsf{T}}\boldsymbol{b}\right\|_{\mathrm{S}}^2 \leq B^2\sigma^2\eta_F^2 s_{\min}(\boldsymbol{E})^{-2}. \qquad (34)$$

Now we can finally replace the bounds (31)–(34) into (30) to get

$$\operatorname{tr}(\operatorname{Var}(\check{\boldsymbol{f}} - \boldsymbol{f})) \leq \sigma^2\left(2(|\boldsymbol{f}|^2 + 1)\eta^2\sum_{i=1}^{m} s_i^{-2} + B^2\eta_F^2\left(2\sum_{i=1}^{m} s_i^{-2} + s_m^{-2}\right)\right).$$

This concludes the main result. For the special case where $\boldsymbol{H}_i' \equiv \boldsymbol{0}$ for all $i$, we have that $\boldsymbol{R}_{\mathrm{NL}} = \boldsymbol{0}$, and then the bounding step in (30) is unnecessary, yielding $\left\|\check{\boldsymbol{f}} - \boldsymbol{f}\right\|_{\mathrm{S}}^2 = \left\|\boldsymbol{E}^{\dagger}(\boldsymbol{R}_w - \boldsymbol{E}_w\boldsymbol{f})\right\|_{\mathrm{S}}^2$. The desired expression is then obtained by merely using (31). ■

## 7.3 Derivation of the SNR equation (14)

In [30], three metrics are suggested to evaluate the quality of the errors calculated by using the first-order approximation. The first two are related to the approximation of the pseudo-inverse, and the last is specific to

least squares:

$$c_1 := \left\|\boldsymbol{E}^\dagger \boldsymbol{E}_w\right\|_{\mathrm{S}}, \tag{35}$$

$$c_2 := \sup_{\boldsymbol{v}\neq \boldsymbol{0}} \mathbb{E}\left[\frac{\boldsymbol{v}^\mathsf{T}\boldsymbol{E}_w^\mathsf{T}\boldsymbol{P}_\perp \boldsymbol{E}_w \boldsymbol{v}}{\boldsymbol{v}^\mathsf{T}\boldsymbol{E}^\mathsf{T}\boldsymbol{E}\boldsymbol{v}}\right], \tag{36}$$

$$c_3 := \frac{\left\|\breve{\boldsymbol{E}}^\dagger - \boldsymbol{E}^\dagger\right\|_{\mathrm{S}}}{\left\|\boldsymbol{E}^\dagger\right\|_{\mathrm{F}}}. \tag{37}$$

Due to the asymptotic arguments that justify the theory in [30], all these metrics must be significantly smaller than 1 for Theorem 3.2 to have practical meaning; i.e., this ensures that the approximation $\|\breve{\boldsymbol{f}} - \boldsymbol{f}\|_{\mathrm{S}} \approx \|\hat{\boldsymbol{f}} - \boldsymbol{f}\|_{\mathrm{S}}$ holds. Below we provide some expressions for upper bounds of these metrics, using the same facts and similar algebraic techniques as in the proof of Theorem 3.2. For brevity, we will keep explanations of steps to a minimum.

$$\begin{aligned} c_1^2 &= \mathbb{E}[\mathrm{tr}(\boldsymbol{E}_w^\mathsf{T}\boldsymbol{E}^{\dagger\mathsf{T}}\boldsymbol{E}^\dagger\boldsymbol{E}_w)] = \mathbb{E}[\mathrm{tr}(\boldsymbol{E}^{\dagger\mathsf{T}}\boldsymbol{E}^\dagger\boldsymbol{E}_w\boldsymbol{E}_w^\mathsf{T})] \overset{\text{Lemma 7.5}}{\leq} \mathbb{E}[\lambda_{\max}(\boldsymbol{E}^{\dagger\mathsf{T}}\boldsymbol{E}^\dagger)\,\mathrm{tr}(\boldsymbol{E}_w\boldsymbol{E}_w^\mathsf{T})] \\ &= \lambda_{\max}((\boldsymbol{E}^\mathsf{T}\boldsymbol{E})^{-1})\,\mathbb{E}[\mathrm{tr}(\boldsymbol{E}_w^\mathsf{T}\boldsymbol{E}_w)] = s_m^{-2}\,\mathbb{E}[\mathrm{tr}(\boldsymbol{E}_w^\mathsf{T}\boldsymbol{E}_w)] \overset{\text{Lemma 7.4}}{\leq} s_m^{-2}\sigma^2 N\gamma_F^2. \end{aligned} \tag{38}$$

For $c_2$, we extract from the proof of Lemma 7.4 that, for any constant vector $\boldsymbol{b} \in \mathbb{R}^m$,

$$\mathbb{E}[\boldsymbol{b}^\mathsf{T}\boldsymbol{E}_w^\mathsf{T}\boldsymbol{E}_w\boldsymbol{b}] \leq \sigma^2\eta_F^2\,|\boldsymbol{b}|^2. \tag{39}$$

Now, define $\boldsymbol{c} := \boldsymbol{E}\boldsymbol{v}$ to get

$$\begin{aligned} c_2 = \sup_{\|\boldsymbol{c}\|=1} \mathbb{E}[\boldsymbol{c}^\mathsf{T}\boldsymbol{E}^{\dagger\mathsf{T}}\boldsymbol{E}_w^\mathsf{T}\boldsymbol{P}_\perp\boldsymbol{E}_w\boldsymbol{E}^\dagger\boldsymbol{c}] &\overset{\boldsymbol{P}_\perp \preceq I}{\leq} \sup_{\|\boldsymbol{c}\|=1} \mathbb{E}[\boldsymbol{c}^\mathsf{T}\boldsymbol{E}^{\dagger\mathsf{T}}\boldsymbol{E}_w^\mathsf{T}\boldsymbol{E}_w\boldsymbol{E}^\dagger\boldsymbol{c}] \\ &\overset{(39)}{\leq} \sup_{\boldsymbol{b}=\boldsymbol{E}^\dagger\boldsymbol{c},\|\boldsymbol{c}\|=1} \sigma^2\eta_F^2\,|\boldsymbol{b}|^2 = \sigma^2\eta_F^2 s_{\max}(\boldsymbol{E}^\dagger)^2 = \sigma^2\eta_F^2 s_{\min}(\boldsymbol{E})^{-2}. \end{aligned} \tag{40}$$

For $c_3$, we have from [30] that $\boldsymbol{E}^\dagger - \breve{\boldsymbol{E}}^\dagger = (\boldsymbol{E}^\mathsf{T}\boldsymbol{E})^{-1}\boldsymbol{E}_w^\mathsf{T}\boldsymbol{P}_\perp - \boldsymbol{E}^\dagger\boldsymbol{E}_w\boldsymbol{E}^\dagger$. We follow operations similar to those in the proof of Lemma 7.4 to get

$$\begin{aligned} \|\boldsymbol{E}^\dagger - \breve{\boldsymbol{E}}^\dagger\|_{\mathrm{S}}^2 &= \|(\boldsymbol{E}^\mathsf{T}\boldsymbol{E})^{-1}\boldsymbol{E}_w^\mathsf{T}\boldsymbol{P}_\perp\|_{\mathrm{S}}^2 + \|\boldsymbol{E}^\dagger\boldsymbol{E}_w\boldsymbol{E}^\dagger\|_{\mathrm{S}} \\ &= \mathbb{E}[\mathrm{tr}(\boldsymbol{P}_\perp\boldsymbol{P}^\mathsf{T}{}_\perp\boldsymbol{E}_w(\boldsymbol{E}^\mathsf{T}\boldsymbol{E})^{-2}\boldsymbol{E}_w^\mathsf{T})] + \mathbb{E}[\mathrm{tr}(\boldsymbol{E}^\dagger\boldsymbol{E}^{\dagger\mathsf{T}}\boldsymbol{E}_w^\mathsf{T}\boldsymbol{E}^{\dagger\mathsf{T}}\boldsymbol{E}^\dagger\boldsymbol{E}_w)] \\ &\leq \mathbb{E}[\mathrm{tr}(\boldsymbol{E}_w(\boldsymbol{E}^\mathsf{T}\boldsymbol{E})^{-2}\boldsymbol{E}_w^\mathsf{T})] + s_m^2\,\mathbb{E}[\mathrm{tr}(\boldsymbol{E}^{\dagger\mathsf{T}}\boldsymbol{E}^\dagger\boldsymbol{E}_w\boldsymbol{E}_w^\mathsf{T})] \\ &\leq 2s_m^{-4}\,\mathbb{E}[\mathrm{tr}(\boldsymbol{E}_w^\mathsf{T}\boldsymbol{E}_w)] \\ &\leq 2s_m^{-4}\sigma^2 N\gamma_F^2, \end{aligned}$$

and, thus,

$$c_3 \leq \frac{\sqrt{2N}s_m^{-2}\sigma\gamma_F}{\sqrt{\sum_{i=1}^m s_m^{-2}}} \leq \frac{\sqrt{2N}s_m^{-2}\sigma\gamma_F}{\sqrt{s_m^{-2}}} = \sqrt{2N}s_m^{-1}\sigma\gamma_F.$$

One may note that bounds satisfy $\sqrt{2}c_1 = c_3$, which are proportional to $\sqrt{N}s_m^{-1}$. The inverse of this quantity, $s_m/\sqrt{N}$, has been discussed in Section 3 as an "effective" singular value of $\boldsymbol{E}$, when considering periodic experiments. Since $c_3$ is the metric specifically built for least-squares problems, we chose $c := c_3$ in Remark 3.4.

## 7.4 Proof of Proposition 4.1

*Proof:* For this proof it suffices to show that for any $i \in \mathbb{N}_{<N}$,

$$\lim_{k\to\infty} s_{\min}(\boldsymbol{E}_{[kN,kN+N-1]}) = \lim_{k\to\infty} s_{\min}(\boldsymbol{E}_{[kN+i,kN+i+N-1]}).$$

Let $\boldsymbol{e}'$ be the limit periodic signal of $\boldsymbol{e}$, i.e., $\boldsymbol{e}'$ is $N$-periodic and, for any $i \in \mathbb{N}_{<N}$, it holds that $\lim_{k\to\infty} \boldsymbol{e}(kN+i) = \boldsymbol{e}'(i)$. We start by noting that

$$\boldsymbol{E}'_{[i,N+i-1]} = \begin{bmatrix} \boldsymbol{E}'_{[i,N-1]} \\ \boldsymbol{E}'_{[0,i-1]} \end{bmatrix} = \begin{bmatrix} \mathbf{0} & \mathbf{I}_i \\ \mathbf{I}_{N-i} & \mathbf{0} \end{bmatrix} \boldsymbol{E}'_{[0,N-1]},$$

where the subscripts denote the dimension of the corresponding identity matrices. Thus,

$$\boldsymbol{E}'^{\top}_{[0,N-1]}\boldsymbol{E}'_{[0,N-1]} = \boldsymbol{E}'^{\top}_{[i,N+i-1]}\boldsymbol{E}'_{[i,N+i-1]},$$

which implies that for all $i$

$$\lim_{k\to\infty} s_{\min}(\boldsymbol{E}_{[kN,kN+N-1]}) = s_{\min}(\boldsymbol{E}'_{[0,N-1]}) = s_{\min}(\boldsymbol{E}'_{[i,N+i-1]}) = \lim_{k\to\infty} s_{\min}(\boldsymbol{E}_{[kN+i,kN+i+N-1]}),$$

proving the existence of the limit. ■

## 7.5 Proofs of Theorem 4.1 and Proposition 4.2

We start with a Lemma for the subgradient information in Theorem 4.1:

**Lemma 7.6 (First-order oracle expression).** *Consider the function $f : \mathbb{R}^n \to \mathbb{R}$ defined by*

$$f(\boldsymbol{x}) := \lambda_{\min}\left(\sum_{i=1}^{N} \boldsymbol{P}_i \boldsymbol{x}\boldsymbol{x}^{\top}\boldsymbol{P}_i^{\top}\right).$$

*Let $\boldsymbol{v}_{\min}$ be a unitary eigenvector associated to one of the minimal eigenvalues of $\sum_{i=1}^{N} \boldsymbol{P}_i\boldsymbol{x}\boldsymbol{x}^{\top}\boldsymbol{P}_i^{\top}$. Then, the vector*

$$\boldsymbol{g} := 2\sum_{i=1}^{N} \boldsymbol{v}_{\min}^{\top}\boldsymbol{P}_i\boldsymbol{x}\boldsymbol{P}_i^{\top}\boldsymbol{v}_{\min}$$

*is a subgradient of $f$.*

*Proof:* The first step is to reformulate the minimum eigenvalue function as the solution of a minimization problem. First, notice that the matrix $\boldsymbol{S}(\boldsymbol{x}) := \sum_{i=1}^{N} \boldsymbol{P}_i\boldsymbol{x}\boldsymbol{x}^{\top}\boldsymbol{P}_i^{\top}$ is symmetric. Therefore, by Rayleigh's Theorem [40, Theorem 4.2.2],

$$f(\boldsymbol{x}) = \min_{|\boldsymbol{v}|_2 \geq 1} \phi(\boldsymbol{v},\boldsymbol{x}) := \min_{|\boldsymbol{v}|_2 = 1} \boldsymbol{v}^{\top}\left(\sum_{i=1}^{N} \boldsymbol{P}_i\boldsymbol{x}\boldsymbol{x}^{\top}\boldsymbol{P}_i^{\top}\right)\boldsymbol{v},$$

for which $\boldsymbol{v}_{\min}$ is in the set of minimizers. Furthermore, the function $\phi(\boldsymbol{v},\boldsymbol{x})$ is convex in both $\boldsymbol{v}$ and $\boldsymbol{x}$. Thus, applying Danskin's Theorem [41], the partial derivative $\partial\phi(\boldsymbol{x},\boldsymbol{v})/\partial\boldsymbol{x}$, taken at $\boldsymbol{v} = \boldsymbol{v}_{\min}$, is a subgradient of $f(\boldsymbol{x})$. Applying the dot product rule gives

$$\frac{\partial\phi(\boldsymbol{v},\boldsymbol{x})}{\partial\boldsymbol{x}} = 2\sum_{i=1}^{N} \boldsymbol{v}^{\top}\boldsymbol{P}_i\boldsymbol{x}\boldsymbol{P}_i^{\top}\boldsymbol{v}.$$

Replacing $\boldsymbol{v}$ for $\boldsymbol{v}_{\min}$ concludes the proof. ■

*Proof of Theorem 4.1:* Since $\mathscr{T}(\mathfrak{z})$ is BIBO stable, by Proposition 4.1 we have that $J_\infty = s_{\min}(\boldsymbol{E}'_{[0,N-1]})^2$, where $\boldsymbol{E}'_{[0,N-1]}$ is the matrix built by the first $N$ samples of the limit periodic signal $\boldsymbol{e}'$.

We now compute $\boldsymbol{e}'$ as a function of $\boldsymbol{u}$. To do this, we must find the initial state $\boldsymbol{x}_0'$ such that, for all $k \in \mathbb{N}_0$, $\boldsymbol{e}'(i+kN) = \boldsymbol{P}_i'\bar{\boldsymbol{u}}$. First, recall (18) and notice that, for any $\boldsymbol{x}_0$,

$$\boldsymbol{x}(N) = \boldsymbol{P}_x\bar{\boldsymbol{u}} + \boldsymbol{A}^N\boldsymbol{x}_0.$$

Now, $\boldsymbol{x}$ is $N$-periodic iff $\boldsymbol{x}(N) = \boldsymbol{x}_0$, hence

$$\boldsymbol{x}_0 = \boldsymbol{x}_0' := (\mathbf{I} - \boldsymbol{A}^N)\boldsymbol{P}_x\bar{\boldsymbol{u}} \tag{41}$$

renders $\boldsymbol{e}$ $N$-periodic and equal, by definition, to $\boldsymbol{e}'$. From LTI systems theory and by definition of $\boldsymbol{P}_i$ and $\bar{\boldsymbol{u}}$, we have that

$$\boldsymbol{e}'(i) = \boldsymbol{P}_i\bar{\boldsymbol{u}} + \boldsymbol{C}\boldsymbol{A}^{i-1}\boldsymbol{x}_0.$$

Substituting $\boldsymbol{x}_0$ for $\boldsymbol{x}_0'$ in (41), and by definition of $\boldsymbol{P}_i'$ in (18), we get

$$\boldsymbol{e}'(i) = \boldsymbol{P}_i'\bar{\boldsymbol{u}},$$

which, due to periodicity of $\boldsymbol{x}$, implies $\boldsymbol{e}'(i+kN) = \boldsymbol{e}'(i),\ \forall k \in \mathbb{N}_0$.

Notice now that $\boldsymbol{E}'_{[0,N-1]} = \begin{bmatrix}\boldsymbol{P}_1'\bar{\boldsymbol{u}} & \boldsymbol{P}_2'\bar{\boldsymbol{u}} & \cdots & \boldsymbol{P}_N'\bar{\boldsymbol{u}}\end{bmatrix}^\mathsf{T}$. For simplicity, denote $\boldsymbol{E}' := \boldsymbol{E}'_{[0,N-1]}$. Now, recall that $J_\infty(\boldsymbol{u}) = s_{\min}(\boldsymbol{E}')^2 = \lambda_{\min}(\boldsymbol{E}'^\mathsf{T}\boldsymbol{E}')$. It is easy to see that

$$\boldsymbol{E}'^\mathsf{T}\boldsymbol{E}' = \sum_{i=1}^{N}\boldsymbol{P}_i'\bar{\boldsymbol{u}}\bar{\boldsymbol{u}}^\mathsf{T}\boldsymbol{P}_i'^\mathsf{T},$$

which establishes that $J_\infty(\boldsymbol{u}) = J(\bar{\boldsymbol{u}})$ as desired.

The subgradient result (19a) is an immediate consequence of Lemma 7.6. ■

*Proof of Proposition 4.2:* First, notice that $\max\left\{\lambda \mid \sum_{i=1}^N \boldsymbol{P}_i'\bar{\boldsymbol{U}}\boldsymbol{P}_i'^\mathsf{T} \succeq \lambda\mathbf{I}\right\} = \lambda_{\min}\left(\sum_{i=1}^N \boldsymbol{P}_i'\bar{\boldsymbol{U}}\boldsymbol{P}_i'^\mathsf{T}\right)$, by the properties of eigenvalues of symmetric matrices. Thus, we may rewrite (20) as the equivalent problem

$$\begin{aligned} \max_{\bar{\boldsymbol{U}}} \quad & \lambda_{\min}\left(\sum_{i=1}^{N}\boldsymbol{P}_i'\bar{\boldsymbol{U}}\boldsymbol{P}_i'^\mathsf{T}\right) \\ \text{subject to} \quad & \operatorname{tr}(\boldsymbol{S}_i\bar{\boldsymbol{U}}) \le 1, \quad \forall i = 1,\ldots,N_c, \\ & \bar{\boldsymbol{U}} \succeq \mathbf{0} \end{aligned} \tag{42}$$

To prove that $\lambda^* \le \lambda^{\mathrm{SDP}}$, it suffices to show that for any feasible $\bar{\boldsymbol{u}} \in \bar{\mathscr{U}}$, $\bar{\boldsymbol{u}}\bar{\boldsymbol{u}}^\mathsf{T}$ is a feasible solution of (42). This is easy to see since $\bar{\boldsymbol{u}}^\mathsf{T}\boldsymbol{S}_i\bar{\boldsymbol{u}} = \operatorname{tr}(\boldsymbol{S}_i\bar{\boldsymbol{u}}^\mathsf{T}\bar{\boldsymbol{u}}) \le 1$, and $\bar{\boldsymbol{u}}\bar{\boldsymbol{u}}^\mathsf{T} \succeq \mathbf{0}$.

To prove equality $\lambda^* = \lambda^{\mathrm{SDP}}$ when $\bar{\boldsymbol{U}}^*$ has rank 1, we can check that in this case $\bar{\boldsymbol{U}}^* = s_1\boldsymbol{v}\boldsymbol{v}^\mathsf{T}$, where $s_1 > 0$ is the only non-zero singular value of $\bar{\boldsymbol{U}}^*$. Then let $\bar{\boldsymbol{u}}^* := \sqrt{s_1}\boldsymbol{v}$, and it holds that $\bar{\boldsymbol{U}}^* = \bar{\boldsymbol{u}}^*\bar{\boldsymbol{u}}^{*\mathsf{T}}$. Now, $\bar{\boldsymbol{u}}^* \in \bar{\mathscr{U}}$, because for all $i$, $\operatorname{tr}(\boldsymbol{S}_i\bar{\boldsymbol{U}}^*) = \operatorname{tr}(\boldsymbol{S}_i\bar{\boldsymbol{u}}^*\bar{\boldsymbol{u}}^{*\mathsf{T}}) = \bar{\boldsymbol{u}}^{*\mathsf{T}}\boldsymbol{S}_i\bar{\boldsymbol{u}}^* \le 1$. Hence, $\lambda_{\min}\left(\sum_{i=1}^N \boldsymbol{P}_i'\bar{\boldsymbol{u}}^*\bar{\boldsymbol{u}}^{*\mathsf{T}}\boldsymbol{P}_i'^\mathsf{T}\right) \le \lambda^* \le \lambda^{\mathrm{SDP}}$. But $\lambda_{\min}\left(\sum_{i=1}^N \boldsymbol{P}_i'\bar{\boldsymbol{u}}^*\bar{\boldsymbol{u}}^{*\mathsf{T}}\boldsymbol{P}_i'^\mathsf{T}\right) = \lambda_{\min}\left(\sum_{i=1}^N \boldsymbol{P}_i'\bar{\boldsymbol{U}}^*\boldsymbol{P}_i'^\mathsf{T}\right) = \lambda^{\mathrm{SDP}}$, hence equality is proven. ■